\documentclass[aps,twocolumn,superscriptaddress]{revtex4}
\usepackage{graphicx}
\usepackage{epsfig}

\usepackage{amsmath}
\usepackage{amssymb}
\usepackage{amsfonts}

\begin{document}
\bibliographystyle{apsrev}
\title{A Solution to the Galactic Foreground Problem for LISA}

\author{Jeff Crowder}
\affiliation{Department of Physics, Montana State University, Bozeman, MT 59717}
\affiliation{Jet Propulsion Laboratory, California Institute of Technology, Pasadena, CA 91109}
\author{Neil J. Cornish}
\affiliation{Department of Physics, Montana State University, Bozeman, MT 59717}

\begin{abstract}

Low frequency gravitational wave detectors, such as the Laser Interferometer Space Antenna (LISA), 
will have to contend with large foregrounds produced by millions of compact galactic binaries
in our galaxy. While these galactic signals are interesting in their own right, the unresolved
component can obscure other sources. The science yield for the LISA mission can be improved if
the brighter and more isolated foreground sources can be identified and regressed from the data.
Since the signals overlap with one another we are faced with a ``cocktail party'' problem of picking out
individual conversations in a crowded room. Here we present and implement an end-to-end solution to
the galactic foreground problem that is able to resolve tens of thousands of sources from across
the LISA band. Our algorithm employs a variant of the Markov Chain Monte Carlo (MCMC) method, which
we call the Blocked Annealed Metropolis-Hastings (BAM) algorithm. Following a description of the
algorithm and its implementation, we give several examples ranging from searches for a single source
to searches for hundreds of overlapping sources. Our examples include data sets from the first round
of Mock LISA Data Challenges. 
\end{abstract}

\maketitle

\section{Introduction}

Galactic compact binary systems are expected to be the major source of gravitational waves
detected by the LISA observatory~\cite{lppa}. Tens of millions of such binaries will be
emitting gravitational waves in the LISA band~\cite{evans,lip,hils,hils2,gils}. Though most
will be have a signal-to-noise ratio (SNR) too low to be detectable, tens of thousands are
expected to be resolvable if optimal signal analysis algorithms are available~\cite{seth}.
The signals from unresolved binary systems will constitute a gravitational wave
confusion noise. Identification of the brighter galactic binaries will be an important aid to
resolving other sources for LISA, such as supermassive black hole binaries
(SMBHBs)~\cite{vech,lang,bbw,rw,kz} and  extreme mass ratio inspirals of compact objects
into supermassive black holes
(EMRIs)~\cite{cutbar,emri,drasco, drasco_hughes}. The SMBHB, EMRI and compact binary signals have
small, but non-vanishing overlap with one another, so we will ultimately want a simultaneous fit
to all sources and source types.

Various techniques have been proposed to extract the parameters of sources from the LISA data stream.
The methods include Markov Chain Monte Carlo Methods~\cite{MCMC, MCMC_ed_neil, vecchio_1, vecchio_2},
genetic algorithms~\cite{genetic}, iterative methods~\cite{gclean, slicedice}, grid-based template
searching~\cite{curt_michele_duncan}, tomographic reconstruction~\cite{browns} and time-frequency
methods~\cite{time_freq}. For most of these methods (tomography and time-frequency analysis
being the exceptions) optimal filtering is accomplished through the
construction of templates describing the signals from all sources in the data stream. For LISA, the
vast number of sources involved makes a direct approach (such as a grid-based template bank)
computationally impractical. It is for this reason ergodic methods such as MCMC and genetic
algorithms have been applied to the LISA data analysis problem. 

In this work we develop an extension of the MCMC method~\cite{metro,haste,gamer} that is able to
search the entire LISA band and simultaneously solve for thousands of galactic binaries.
Our approach is based on the observation that while some signals can have significant overlap,
signals that are well separated in frequency have little or no overlap. We exploit this
quasi-locality by breaking the search up into sub-regions in frequency, taking care with
edge effects. In a departure from our previous approach~\cite{MCMC}, we do not try and
update all the source parameters simultaneously at each iteration. This greatly reduces
the computational cost, while still providing a global solution as the full meta-template
comprising all the search templates is used to evaluate the likelihood. The parameter
updates are now done in small blocks of highly correlated parameters, and the solution
is updated using Metropolis-Hastings sampling. Simulated annealing is employed during
the search phase to improve the mixing of the chains. We demonstrate this ``Blocked
Annealed Metropolis-Hastings'' (BAM) algorithm on simulated LISA data
that contains the signals from monochromatic white dwarf binary systems (WD-WD) immersed in
Gaussian instrumental noise. 

The paper is organized as follows: In Section~\ref{algorithm_overview} we review
the MCMC algorithm and the Metropolis-Hastings sampling kernel. Additionally, we give a description
of the BAM algorithm and demonstrate how its computational cost scales with the number of sources being
searched for. Section~\ref{algorithm_issues} discusses various aspects of using the BAM algorithm,
such as hierarchical searching, optimal choices for search ranges, stopping criteria, and rates of
occurrence for false positives and false negatives. Example searches are performed in
Section~\ref{example_searches}, ranging from individual sources to hundreds of sources in a
restricted range of the LISA band. Concluding remarks are made in Section~\ref{conclusion}.

\section{Algorithm Overview}\label{algorithm_overview}

In this section we describe the BAM algorithm and the issues and difficulties that we worked through in the
development process. We start with a brief review of the basic MCMC approach and Metropolis-Hastings
importance sampling. This is followed by a review of how the extrinsic source parameters are
analytically solved for using the F-Statistic. We then describe how the speed of the BAM
algorithm scales with the number of sources that are being searched for.

\subsection{The Markov Chain Monte Carlo Algorithm and Metropolis-Hastings Sampling}

The MCMC algorithm is becoming a familiar tool in gravitational wave data analysis. Initially introduced
to the field by Christensen and Meyer~\cite{cm1}, its application to ground-based interferometers
has been explored in the context of parameter extraction of coalescing binaries~\cite{christ} and spinning neutron stars~\cite{woan}. With its ability to explore large
parameter spaces while simultaneously performing model selection and noise estimation, the MCMC method
is ideally suited to the LISA data analysis problem. The Reverse Jump MCMC algorithm has
been applied to the LISA-like problem of identifying a large, yet unknown number of sinusoids in simulated
Gaussian noise~\cite{andrieu, woan2}. It was shown that the method could correctly identify the
number of resolvable signals present in the data and recover the signal parameters and an estimate
of the noise level. The MCMC approach was first applied to simulated LISA data in the context of
galactic binaries~\cite{MCMC}, and it has since been applied to SMBHBs~\cite{MCMC_ed_neil, vecchio_1, vecchio_2}.

In using an MCMC approach one wants to generate a sample set, $\{ \vec{x} \}$ that corresponds to draws
made from the posterior distribution of the system, $p(\vec{\lambda} \vert s)$. The algorithm to develop
such a set is surprisingly simple.

We begin at a point in the parameter space of the binary
system(s), $\vec{x}$, (which may or may not be chosen at random) and propose a jump
to a new position, $\vec{y}$, based on some proposal
distribution, $q(\cdot \vert \vec{x})$. The Hastings ratio is calculated using,
\begin{equation}\label{Hastings_ratio}
H = \frac{p(\vec{y}) p(s \vert \vec{y}) q(\vec{x} \vert \vec{y})}
{p(\vec{x}) p(s \vert \vec{x}) q(\vec{y} \vert \vec{x})} \, ,
\end{equation}
where $p(\vec{x})$ is the prior of the parameters at $\vec{x}$, $q(\vec{x} \vert \vec{y})$ is the value of
the proposal distribution for a jump from $\vec{x}$ to $\vec{y}$, and $p(s \vert \vec{x})$ is the likelihood
at $\vec{x}$. The likelihood function, if the noise is a normal process with zero mean, is given by~\cite{sam}
\begin{equation}\label{likely}
p(s \vert \vec{x}) = C \exp\left[ -\frac{1}{2} \left( (s - h(\vec{x})) \vert (s - h(\vec{x}))
\right) \right]\, ,
\end{equation}
where the normalization constant $C$ is independent of the signal, $s$, and $(a \vert b )$ denotes the
noise weighted inner product
\begin{equation}\label{general_inner_product}
(a \vert b ) = 2 \int_0^\infty \frac{ \tilde{a}^{\ast}(f) \tilde{b}(f) + \tilde{a}(f) \tilde{b}^{\ast}(f)}{S_n(f)} \,df ,
\end{equation}
where $a$ and $b$ are the gravitational waveforms, and $S_n(f)$ is the one-sided noise spectral density. 
The jump will be accepted with probability $\alpha = {\rm min}(1,H)$. If the jump is rejected
(Metropolis rejection~\cite{metro}) the chain remains at its current state, $\vec{x}$. Repeated jumps
will produce a Markov chain whose stationary distribution is equal to the posterior
distribution in question, $p(\vec{x} \vert s)$. Andrieu {\it et al}~\cite{mcmc_hist, andrieu}
provide a more general and thorough review of MCMC methods.

The convergence to the correct posterior will occur for any (non-trivial) proposal distribution~\cite{gilks}.
However, the more accurate the proposal distribution is at modeling the posterior distribution the quicker
the chain will convergence. Since we do not know the form of the posterior in advance of running the
chain, we instead opt for maximum flexibility in our choice of proposal distribution. We accomplish this
by using a mixture of proposal distributions, including occasional ``bold'' proposals that attempt
large changes in the parameter values along with many ``timid'' proposals that attempt small changes in
the parameter values of the chain (for a detailed description of some of these proposals see~\cite{MCMC}).
Also added to our list of proposals are a few ``tailored'' proposals. These are proposed jumps based
on our knowledge of the symmetries and degeneracies of the likelihood surface. For example, for LISA
there is a secondary maxima in the likelihood surface of a binary system that corresponds to a reflection
about the ecliptic equator combined with a shift of $\pi$ in the ecliptic longitude. Thus, we include
a proposal that attempts such a reflection.

\begin{figure}[h]
\includegraphics[angle=270,width=0.5\textwidth]{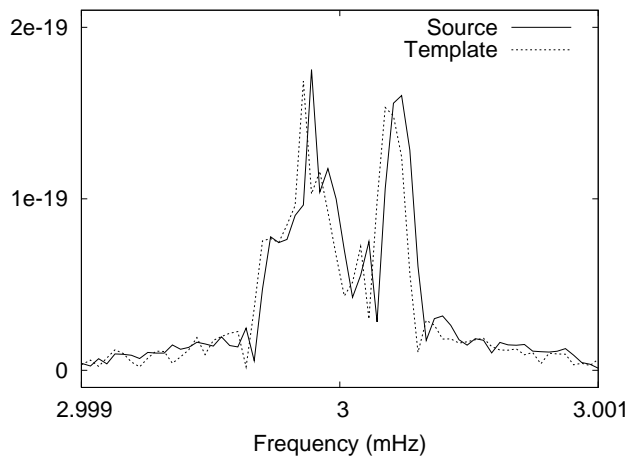}
\caption{\label{fm_jump_figure_1}Spectrum of a source and template with the same parameters
save their frequencies, which
differ by one modulation frequency. The source has $f = 3.0 {\rm mHz}$, $A = 1.4 \times 10^{-22}$,
$\theta = 1.0$, $\phi = 3.14159$, $\psi = 0.5$, $\iota = 0.785398$, and $\varphi_o = 0.0$.}
\end{figure}

\begin{figure}[h]
\includegraphics[angle=270,width=0.5\textwidth]{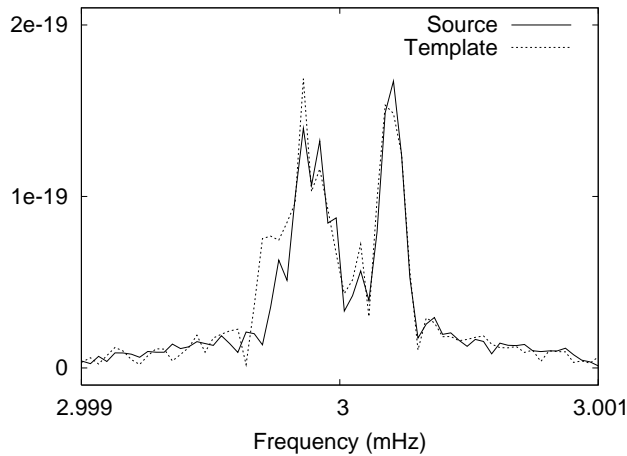}
\caption{\label{fm_jump_figure_2}Spectrum of a sources and a template with frequencies that
differ by one modulation
frequency. The parameters of the template have been adjusted to maximize the overlap. The source
has the same parameters as in Figure 1, while the template has $A = 1.47254 \times 10^{-22}$, $f = 3.000031688 {\rm mHz}$, $\theta = 0.787087$, $\phi = 3.18781$, $\psi = 0.51218$, $\iota = 0.89419$, and $\varphi_o = 0.179034$.}
\end{figure}

Another tailored proposal that was highly effective uses the fact that local maxima occur in the
likelihood surface at multiples of the modulation frequency ($f_m = 1/{\rm year}$). The presence of
these maxima are easily understood. The detector response imparts sidebands on the monochromatic
Barycenter signal to produce a comb whose teeth are spaced by the modulation frequency. A template
with a Barycenter frequency offset from the signal by a multiple of $f_m$ will produce a similar
comb whose teeth align with those of the signal. The fit can be improved by adjusting the other
template parameters to better match the shape of the source comb. Figure~\ref{fm_jump_figure_1} shows
the power spectrum of a source and a template offset by $1 f_m$ in frequency, while in 
Figure~\ref{fm_jump_figure_2} the template has had its other parameters altered to better fit
the source comb. This improves the log likelihood from $16.7\%$ to $66.4\%$ of the true parameter
value. It proved to be difficult to find an analytic description of how the other parameters
should be modified to improve the fit, so we went with a simple proposal that shifted the frequency
of the source by $1 f_m + \epsilon$, and shifted the other parameters using a uniform draw in a
small range around the current binary parameters.

Since the search used an F-statistic to extremize over the amplitude, inclination angle, polarization
angle and initial phase, we only needed to assign priors to the frequency and sky location. For the
frequency we used a uniform prior across the search range, and for the sky locations we used a
distance weighted galactic distribution.
 
\begin{figure}[h]
\includegraphics[angle=0,width=0.5\textwidth]{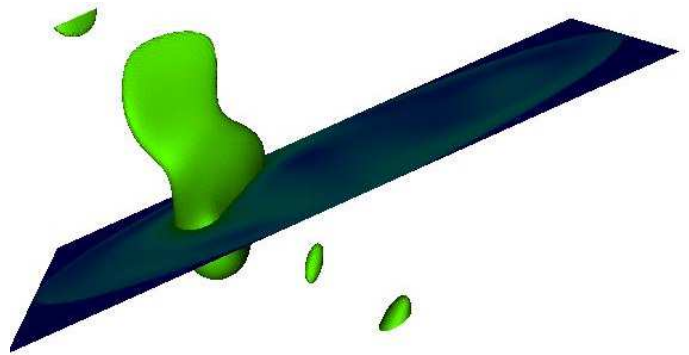}
\caption{\label{island_chain}Contours of constant likelihood for a single, monochromatic WD-WD binary system.
Sky position directions lie in the plane shown, while frequency is directed orthogonal to the plane.
The visible maxima are separated by approximately one modulation frequency.}
\end{figure}

While a general MCMC algorithm functions as both a means to search the likelihood surface and sample from
the posterior distribution, the Markovian nature is only required for the sampling portion of the process.
If one is able to find the neighborhood of the true parameters by other means, and then
implements an MCMC algorithm from that point, the subsequent samples will recover the
correct posterior distribution.
It is with this in mind that we chose to relax the requirement that all of our proposal distributions be
Markovian. The algorithm described above remains the same, but is now open to non-Markovian proposal
distributions. We refer to this non-MCMC approach as Metropolis-Hastings importance sampling. One example
of a non-Markovian proposal involves how we implement the previously described $1 f_m$ jump. Early in the
search phase, if one of these proposals was accepted, a second (identical) jump was attempted. The reason
for this is that just as a local maximum occurs $\sim 1 f_m$ away from the global maximum, another (smaller)
local maximum occurs $\sim 2 f_m$ away. In fact, a chain of maxima occur spaced about one modulation frequency
apart from each other in likelihood space (see Figure~\ref{island_chain}). This non-Markovian move allowed
searchers to ``island-hop'' around the likelihood surface if it found itself on one of the maxima in the
island chain.

For the purpose of searching to determine the neighborhood of the true parameters, which is the focus of
this work, we allow ourselves the extra freedom provided by non-Markovian Metropolis-Hastings importance
sampling and simulated annealing (described below). Once the search phase is complete the non-Markovian
moves are suspended, and the algorithm performs a standard MCMC exploration of the posterior distribution.

To encourage the chains to explore the full parameter space and to discourage the chains from
getting stuck a local maxima we employ simulated annealing during the search phase. This is done
by multiplying the noise-weighted inner product by an ``inverse temperature'' $\beta$, and
applying the cooling schedule
\begin{equation}
\beta = \left\{
\begin{tabular}{ll}
$\beta_0 \left(\frac{1}{\beta_0}\right)^{i/N_c}$ & $0 \leq i \leq N_c$ \\
 1 &  $i > N_c $
\end{tabular} .
\right.
\end{equation}
Here $\beta_0$ is the initial heating factor and $N_c$ is the length of the annealing phase. The
choice of $\beta_0$ depends on the SNR of the sources. When very bright sources are present we
set $\beta_0 \sim 10^{-3}$, while smaller values of $\beta_0 \sim 10^{-2}$ work better once the
brightest sources have been removed. Our criteria for choosing $\beta_0$ was that the chains
should explore the full parameter range during the first 1000 or so iterations. If full range
movement was not seen we dialed up the heat. The choice of $N_c$ depends on $\beta_0$, and
can be set by demanding that it takes, say, 500 steps for $\beta$ to decrease by a factor of two.

\subsection{F-Statistic}

The F-statistic~\cite{fstat} uses multiple linear filters to obtain the extremum for the likelihood
over the extrinsic parameters of the signal. Using the F-statistic one can search the intrinsic parameters,
and recover the extrinsic parameters as a last step in the process. For a monochromatic binary detected by
LISA, this reduces the search space to three parameters: frequency and sky location ($\theta$, $\phi$).
Note: the tumbling orbit of LISA induces modulations of the frequency into the signal, thus creating and
interdependency in the signal between frequency and sky location that prevents $\theta$ and $\phi$ from
being treated as purely extrinsic quantities. 

At low-frequencies in the LISA response to a gravitational wave can be written:
\begin{equation}\label{grav_wave}
h(t) = h_+(t) F^+(t) +  h_\times(t) F^\times(t) \, ,
\end{equation}
where $h_+(t)$ and $h_\times(t)$ are the two polarizations of the incident gravitational wave~\cite{cc,cr},
which for a monochromatic binary, to leading post-Newtonian order, are given by:
\begin{eqnarray}
h_+(t) &=& A (1+\cos^2\iota) \cos(\Phi(t)+\varphi_0) \nonumber \\
h_\times(t) &=& -2 A \cos\iota\, \sin(\Phi(t)+\varphi_0).
\end{eqnarray}
$F^+(t)$ and $F^\times(t)$ are the beam pattern factors:
\begin{eqnarray}
F^+(t) &=& \frac{1}{2} \left( \cos 2 \psi \, D^+(t) - \sin 2 \psi \, D^\times(t)\right) \nonumber \\
F^\times(t) &=& \frac{1}{2} \left( \sin 2 \psi \, D^+(t) + \cos 2 \psi \, D^\times(t)\right)
\end{eqnarray}
where $D^+(t)$ and $D^\times(t)$ are the detector pattern functions, whose mathematical form can be
found in equations (36) and (37) of Ref.\cite{rigad}. 

In the gravitational wave phase
\begin{equation}
\Phi(t; f, \theta, \phi) = 2\pi f t + 2 \pi f {\rm AU} \sin\theta \cos(2 \pi f_m t - \phi),
\end{equation}
one can see the coupling of the sky location and the frequency through the term that depends on the
radius of LISA's orbit, 1 AU, and its orbital modulation frequency, $f_m = 1/{\rm year}$. For the
low frequency galactic sources we are considering, the gravitational wave amplitude, $A$, is
effectively constant. Thus (\ref{grav_wave}) can be rearranged as:
\begin{equation}\label{grave_wave_sum}
h(t) = \sum_{i=1}^4 a_i(A,\psi,\iota,\varphi_0) A^i(t; f, \theta, \phi)\, ,
\end{equation}
where the time-independent amplitudes $a_i$ are given by
\begin{eqnarray}
a_1 &=& \frac{A}{2} \left( (1+\cos^2\iota)\cos\varphi_0 \cos 2\psi-2\cos\iota \sin\varphi_0 \sin 2 \psi \right),
\nonumber \\
a_2 &=& -\frac{A}{2} \left( 2\cos\iota\sin\varphi_0 \cos 2\psi+ (1+\cos^2\iota) \cos\varphi_0 \sin 2 \psi \right),
\nonumber \\
a_3 &=& -\frac{A}{2} \left( 2\cos\iota\cos\varphi_0 \sin 2\psi+ (1+\cos^2\iota) \sin\varphi_0 \cos 2 \psi \right),
\nonumber \\
a_4 &=& \frac{A}{2} \left( (1+\cos^2\iota)\sin\varphi_0 \sin 2\psi-2\cos\iota \cos\varphi_0 \cos 2 \psi \right),
\nonumber \\
\end{eqnarray}
and the time-dependent functions $A^i(t)$ are given by
\begin{eqnarray}
A^1(t) &=& D^+(t;\theta, \phi)  \cos \Phi(t;f, \theta, \phi)
\nonumber \\
A^2(t) &=& D^\times(t;\theta, \phi) 
\cos \Phi(t;f, \theta, \phi) \nonumber \\
A^3(t) &=& D^+(t;\theta, \phi) 
\sin \Phi(t;f, \theta, \phi) \nonumber \\
A^4(t) &=& D^\times(t;\theta, \phi) 
\sin \Phi(t;f,\theta, \phi) \, .
\end{eqnarray}
Writing LISA's signal as a superposition of gravitational waves and noise,
\begin{equation}
s_\alpha(t) = h_\alpha(t, \vec{\lambda}) + n_\alpha(t) = \sum_{i=1}^N h^i_\alpha(t,\vec{\lambda}_i)+ n_\alpha(t) \, ,
\end{equation}
and defining the four constants $N^i=(s \vert A^i)$ and the $M$-matrix, $M^{ij}=(A^i\vert A^j)$, one can express (\ref{grave_wave_sum}) as a matrix equation:
\begin{equation}\label{M_a_N_equation}
M_{ij} a^j = N_i\, .
\end{equation}
Therefore, with a signal from LISA and the three intrinsic parameter values, one can solve
(by iteration or inversion) for the the amplitudes, and thus the extrinsic parameters.
The values of the intrinsic parameters are found by use of iterative Metropolis-Hastings importance
sampling. The extrinsic parameter values are given by: 
\begin{eqnarray}
A & = & \frac{ A_+ + \sqrt{A_+^2-A_\times^2}}{2} \nonumber \\
\psi & =& \frac{1}{2}\arctan\left(\frac{A_+ a_4 - A_\times a_1}{-(A_\times a_2 + A_+ a_3)}\right) \nonumber \\
\iota & = & \arccos\left(\frac{-A_\times}{A_+ + \sqrt{A_+^2-A_\times^2}}\right)\nonumber \\
\varphi_0 & =& \arctan\left(\frac{c(A_+ a_4 - A_\times a_1)}{-c(A_\times a_3 + A_+ a_2)}\right)
\end{eqnarray}
where
\begin{eqnarray}
A_+ &=& \sqrt{(a_1+a_4)^2 + (a_2-a_3)^2} \nonumber \\
&& + \sqrt{(a_1-a_4)^2 + (a_2+a_3)^2} \nonumber \\
A_\times &=& \sqrt{(a_1+a_4)^2 + (a_2-a_3)^2} \nonumber \\
&& - \sqrt{(a_1-a_4)^2 + (a_2+a_3)^2} \nonumber \\
c &=& {\rm sign}(\sin(2 \psi)) \, .
\end{eqnarray}
This description of the F-statistic automatically incorporates the two independent LISA channels
through the use of the dual-channel noise weighted inner product: 

We generalize the F-statistic to handle $N$ overlapping sources by writing $i= 4K +l$, where $K$ labels
the source and $l = 1 \rightarrow 4$ labels the four filters for each source. The F-statistic has the
same form, but with $4N$ linear filters $N^i$, and $M^{ij}$ is a $4N \times 4N$ dimensional matrix.
For slowly evolving galactic binaries, which dominate the confusion problem, the limited bandwidth
of each individual signal means that the $M^{ij}$ is band diagonal, which lessens the difficulty in
solving (\ref{M_a_N_equation}) for the large numbers of sources that are expected to be detected.

\subsection{The Blocked Annealed Metropolis-Hastings Algorithm}\label{BAM}

A Gibbs Sampler~\cite{gemans} is a special case of the Metropolis-Hastings algorithm described
above in which each parameter, in the set of all parameters, is updated sequentially using a
proposal distribution that is the conditional of the posterior evaluated at the current values
of the other parameters. The mixing rate can be improved by simultaneously updating blocks
of highly correlated parameters, to give what is know as the Blocked-Gibbs algorithm~\cite{jensen}.
Since it can be difficult to evaluate the full conditionals required for Gibbs sampling, we
decided to stick to Metropolis-Hastings sampling.

\begin{figure}[h]
\includegraphics[angle=270,width=0.5\textwidth]{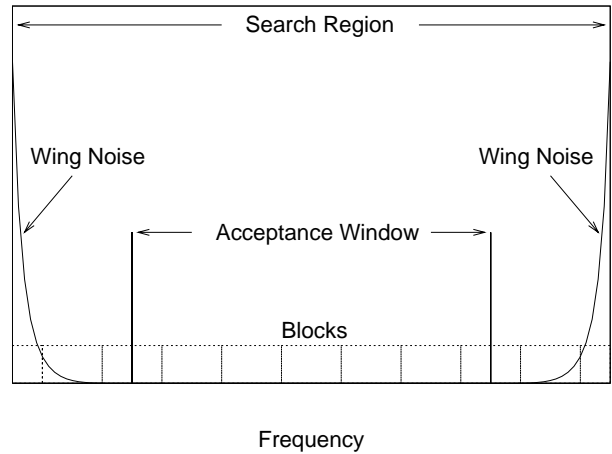}
\caption{\label{Block_MH_figure}A schematic representation of a search region in frequency space, showing
the block structure of some the BAM algorithm.}
\end{figure}

The blocks in the BAM algorithm are small sub-units of the frequency range being searched. As can be
seen in Figure~\ref{Block_MH_figure}, which shows a schematic representation of a search region in our
BAM algorithm, the search region is broken up into equal sized blocks. The algorithm steps through
these blocks sequentially, updating all sources within a given block simultaneously. After all blocks
have been updated, they are shifted by one-half the width of a block for the next round of updates.
This allows two correlated sources that might happen to be located on opposite sides of a border
between two neighboring blocks to be updated together on every other update. This blocking provides
a means to handle highly correlated searchers/sources, and provides an even greater benefit which is
covered in ~\ref{time_scaling}.

This simple extension to an MCMC algorithm allowed for quick and robust searching of isolated data snippets
with up to $\sim 100$ templates (the limit on the number of templates is due to the
computational cost of the multi-source F-statistic). In order to be able to handle the entire LISA band,
we have to break the search up into sub-regions containing a manageable number of sources.
This introduced the problem of edge effects from sources whose frequency lay just outside
the chosen search region, but deposited power into the search region. To combat the edge effects
we introduced ``wings'' and an ``acceptance window.'' The purpose of the wings is to create a buffer between the chosen
search region and the regions beyond, so that sources in those outer regions would not adversely affect
the searchers in the acceptance window. The search would be run the same as before, but at the end only
those searchers inside the acceptance window would be considered as having been found. Searchers that ended
up in the wings were discarded (even though many might be perfectly good fits to actual sources). This would
allow us step through frequency space, using multiple search regions. But that was not enough.

\begin{figure}[h]
\includegraphics[angle=270,width=0.5\textwidth]{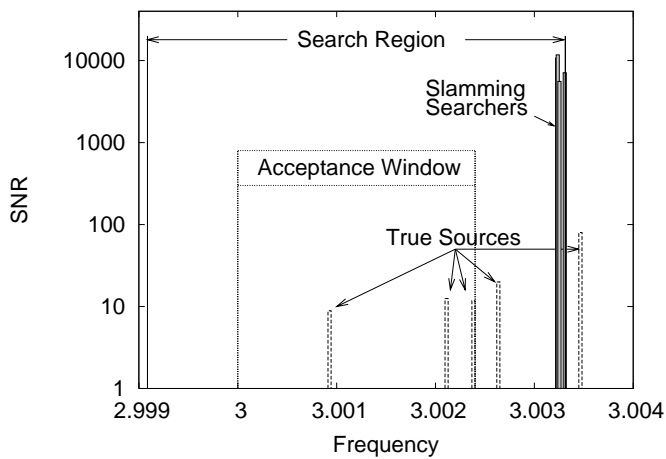}
\caption{\label{slamming_figure}An example of slamming, in which four searchers are drawn to the edge of the search region by a bright source placed just outside the search region.}
\end{figure}

A problem, which we call ``slamming,'' can occur when a bright source lies just outside the range of the
chosen search region. Slamming is the tendency of the searchers to migrate to the edge of the search region
as the searchers try to fit to the portion of the power from the source outside the search region which is
bleeding into the search region. Since the search region contains only a portion of the signal from that
particular source, a single searcher is a poor match, and so other searchers soon are recruited to fix-up
the fit. Figure~\ref{slamming_figure} shows a case of slamming, where four searchers (in a data snippet with
four sources) are drawn off to the edge of the search region by a bright source just outside the search
region. A tell-tale feature of slamming is the large amplitudes of the searchers and their high degree of
correlation or anti-correlation.

\begin{figure}[h]
\includegraphics[angle=270,width=0.5\textwidth]{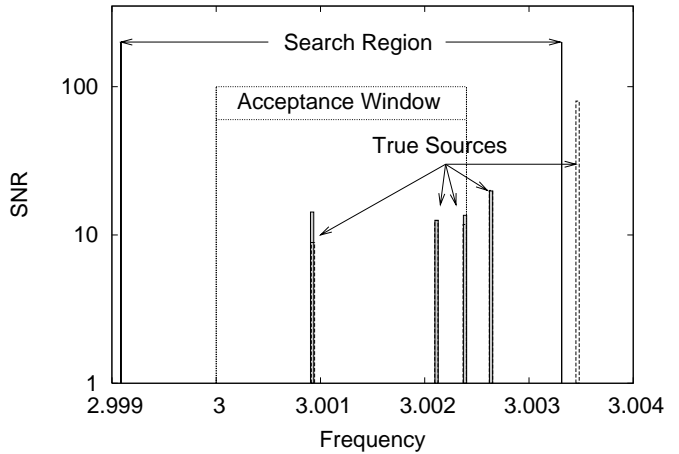}
\caption{\label{non_slamming_figure}The same search shown in Figure~\ref{slamming_figure}, but with the inclusion of wing noise. The wing noise has cured the slamming.}
\end{figure}

One fix to the slamming problem is to weight the matches in the wings less than matches in the acceptance
window. We do this by attenuating the contribution from the wings in the noise weighted inner product.
The attenuation is done by increasing the noise spectral density in the wings, which we call ``wing noise.''
We exponentially increase the noise spectral density, starting at the edge of the acceptance window, as
shown in Figure~\ref{Block_MH_figure}. This addition to the algorithm is quite effective at lessening
the impact on the likelihood of a searcher fitting power bleeding into wings of the search region.
Figure~\ref{non_slamming_figure} shows a search using wing noise in the same region as
Figure~\ref{slamming_figure}. As can be seen there is no slamming, and the searchers were able to find
the true sources.

Another fix to the slamming problem is to use information about bright sources that lay outside
the search region, but close enough that their signals overlap with the search region, in the search
process (e.g. subtracting the signal of such sources from the data snippet). This requires knowledge
of the bright sources, but such knowledge can be gained by performing initial, exploratory searches for
a few sources in the area surrounding the search region in question. As one is free to choose the number
of searchers and the ranges of the search, one can use the BAM algorithm hierarchically, if desired, to
seek out the brightest sources in a pilot search of the data. Information gleaned from the pilot searches
can then be used to subtract bright sources that might lead to slamming. This hierarchical approach will
be described in more detail in the Section~\ref{hierarchical_method}. In general use of our BAM algorithm,
we use both wing noise and information about bright sources outside the search regions to lessen the
chance of slamming. 

\subsection{Time Scaling}\label{time_scaling}

Since there are expected to be $15,000^+$ resolvable galactic binary systems in the LISA data, we need to
be sure that any algorithm developed will be able to search for such a large number of sources in a practical
period of time. A pure BAM algorithm that employed a meta template that is the sum of templates from
individual sources and simple proposal updates would have a computational cost that scaled linearly in the
the number of sources since the cost per source would be constant. The goal of a purely linear scaling is
unrealizable in the current version of the BAM algorithm for two reasons. 

First, in using the F-statistic to calculate the likelihood values we are required to solve equation
(\ref{M_a_N_equation}) for the time-independent amplitudes $a_i$. Inverting this equation would scale
as $N^3$, and while solving for the amplitudes using an iterative method is significantly better it
does not scale linearly with $N$. The next update to the algorithm will be to discontinue use of the
F-statistic and return to a seven parameter (per binary system) search.

\begin{figure}[h]
\includegraphics[angle=270,width=0.5\textwidth]{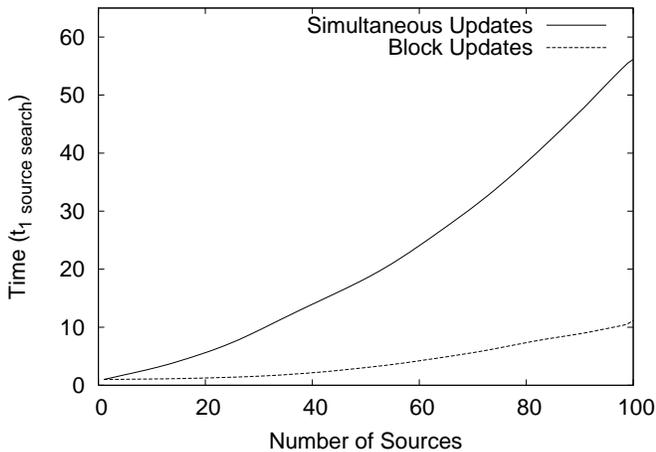}
\caption{\label{time_scaling_figure}Plot of the average computational time per update per search template
for two cases, one where all sources are updated simultaneously, and the other where block updates
are used. The computational time (y-axis) has been scaled by the average time per update for a single
source search. The increase in cost per template update grows much more slowly for the blocked search.
To achieve linear scaling in the total cost of the BAM algorithm
we would need a constant computational cost per individual template update.}
\end{figure}

The second impediment is that some proposal distributions, such as a multivariate normal distribution
jumping along the eigendirections of the variance-covariance matrix are inherently non-linear in
computational cost (calculation of the Fisher Information Matrix alone scales as $N^2$). It is here that
the use of the blocks shows great benefit. Figure~\ref{time_scaling_figure} shows how the time to perform
a searcher update is affected by the number of searchers. The plot shows the average cost per individual
searcher that is updated in a search using only multivariate normal proposal distributions. In one case,
all searchers are updated simultaneously (i.e. the entire search region is treated as one block). In
the other case, breaking the updates into the blocks lessens the number of searchers being updated
simultaneously, greatly reducing the overall computational cost as the number of sources grows. The
search was performed on a one year data stream in a $0.01 {\rm mHz}$ snippet starting at frequency
of $3 {\rm mHz}$ with blocks that were $4/year$ wide. Up through $N=20$, no blocks contained more than
one search template and the increase in computational cost per search template update was just $1.25$
times greater than a single template search, due solely to the use of the multi-source F-statistic.
Beyond $N=20$ the blocks began to contain more than one searcher, and the non-linear nature of the
normal proposal started to drive up the cost per update.

\section{Issues when Using the Algorithm}\label{algorithm_issues}

Thus far we have described the features of the BAM algorithm, and discussed how those features aid in
optimizing the algorithm. In this section some of the implementation issues will be addressed. We will
discuss choices of the sizes of search regions, the sizes of the wings, hierarchical searching,
stopping criteria, and false-positive/false-negative levels.

\subsection{Hierarchical Method}\label{hierarchical_method}

We are able to perform searches in a hierarchical manner simply by choosing to use less search templates
than their are sources. For example, if we search a region that contains $10$ sources with a meta-template
built from $2$ galactic binary templates,
the algorithm will generally recover the $2$ sources that return the highest likelihood value (usually
they will correspond to the highest SNR sources in the region). In this hierarchical approach the solution
for the brightest sources will be thrown off by the sources that were neglected, but this bias can
be corrected at subsequent stages in the hierarchy. At the next stage in the hierarchy more galactic binary
templates are used, with the information from the earlier searches providing starting locations for
some of the search templates.

\begin{figure}[h]
\includegraphics[angle=270,width=0.5\textwidth]{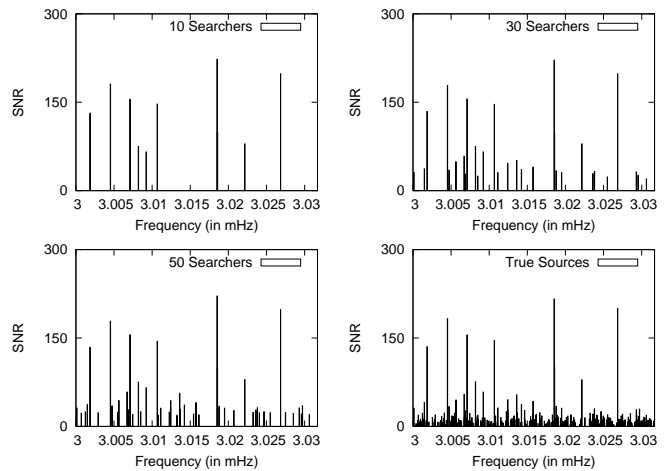}
\caption{\label{hierarchical_example}Plot of the recovered (and true) parameter frequencies and
their respective SNRs for the sources recovered by the BAM algorithm searching a $1000 f_m$ data
snippet containing $281$ sources with a $1$ year time of observation using $10$, $30$, and $50$ searchers.}
\end{figure}

Figure~\ref{hierarchical_example} describes an example where a $1000 f_m$ data snippet containing
$281$ sources with a observation time of one year was searched using $10$, $30$, and $50$ searchers.
In the case using $10$ searchers, the $10$ sources with the highest SNRs were found, while for the case
with $30$ searchers only $25$ of the $30$ highest SNR sources were among those found (the other $5$
searchers did find true sources, albeit with lower SNRs). 

One may be tempted to subtract the signals of sources found in the pilot search, before looking for
more sources in the region, but our experience leads us to believe that this is not the best way to
proceed. In a hierarchical search the initial solution for the bright sources will be thrown off by
dimmer sources that they overlap with. To stop these errors from propagating it is better to allow
the algorithm to refine the fit to the bright sources at subsequent stages in the hierarchy.

Instead of removing the signal from the data stream, we include the information in the meta-template
represented by the combination of the search templates. Within the framework of our algorithm one has
several options of how to include the information at subsequent stages. One option is to begin the next
search with some of the search templates at the source locations determined at the previous
iteration, then allowing them to behave like any other searchers from that point on. Another
option is to hold the intrinsic parameters of the previously determined signals fixed during the
high heat portion of the annealing phase (the extrinsic parameters of all templates are still updated
since we are using a multi-template F-Statistic). The advantage of the later approach is that it
protects the information gleaned from earlier stages in the hierarchy, as there is a danger that
a previously determined template will be dislodged during the annealing phase. When the temperature
reaches a selected value, the constraints are withdrawn and the inherited search templates get
treated like any other. This allows the information gathered in the earlier runs to be preserved,
and also allows for adjustments to be made to the parameters as needed due to the inclusion of
other searchers.

Another benefit is that such hierarchical searching can be used to improve efficiency. Knowing where
the very bright sources are in the data snippet can be used to lessen the sizes of the wings of the
search regions. This will be discussed in more detail in the next subsection.

\subsection{Choosing Window Sizes}

There are several factors that influence the choice of window size. In Subsection~\ref{time_scaling} we saw
that the cost per template update grows with the number of templates in a window. This argues for using
small search windows. However, in Subsection~\ref{BAM} we saw that it was necessary to include a wing
region to mitigate edge effects, and any signals recovered in the wings had to be discarded. Thus, we
would like for the wings to cover a small fraction of the search window, so that the fraction of templates
assigned to wings is small. The size of the wings is determined by the typical bandwidth of a source. A
natural choice is to set the wing size equal to the typical half-bandwidth as this ensures that
little power leaks into the acceptance window. The source bandwidth is determined by two factors, the
width of the sideband comb:
\begin{equation}
B_c \approx 2 \left(5 + 2 \pi f \frac{1 {\rm AU}}{c} \right) f_m
\end{equation}
and the degree of spectral leakage caused by the finite observation time. The spectral leakage
depends on the amount the carrier frequency differs from being an integer multiple of $1/T_{\rm obs}$,
and falls off inversely with the number of frequency bins. For observation times on the order of
years and for frequencies in the milli-Hertz range, the two contributions to the bandwidth are
of similar size. Somewhat larger wings are needed if very bright sources have not been regressed
from neighboring frequency windows.

These physical considerations dictate a total wing size of $\sim 15 \rightarrow 30 f_m$, and we
would like to use snippets considerably larger than this in order to maximize the fraction of
the templates that appear in the acceptance window. The problem with this is that in regions of
high source density we end up with template densities of 1 per $\sim 4$ frequency bins, so for
a 1 year data stream the wings alone account for $4 \rightarrow 7$ templates. Thus, to keep us
in the regime where the cost per update is close to that of a single source, we were forced to
use snippets where the acceptance window was comparable in size to the wings. In future upgrades
we hope to improve the scaling of the algorithm so that we can use larger search windows.

\subsection{Stopping Criteria}

In LISA data analysis we will not only have to determine source parameters, but also the number
of sources that can be resolved. Studies of the galactic confusion noise
levels~\cite{hils,hils2,gils,sterl,seth,bc} provide some answers concerning source populations
across the LISA band, as well as estimates of the number of resolvable binaries as a function
of frequency~\cite{seth}. However, this information is better suited to determining the number of
source templates to start with, than the number with which to end. In order to discover where we
must stop we will have to listen to the data.

Models with more parameters will always produce better fits to the data, but beyond a certain point
the recovered parameters become meaningless. What we seek is the best fit to the data with the
simplest possible model. For a model with uniform priors we seek to minimize
\begin{equation}
\chi^2 = (s-h\vert s-h)\, ,
\end{equation}
using the smallest number of source templates. Using Fisher matrix techniques it can be shown
that the expectation value of this quantity is given by
\begin{equation}
\langle \chi^2 \rangle = {\cal N} - D\, ,
\end{equation}
where ${\cal N}$ is the total number of data points and $D$ is the model dimension. As one might
have anticipated, it does not make sense to use a model with more parameters than there are data
points. This sets an upper limit of 1 source template every $\sim 2$ frequency bins, as there are
4 data points per frequency bin (2 independent data channels, each with a real and
imaginary part), and 7 parameters per template. More refined criteria, such as the
Bayesian evidence, set more stringent stopping criteria.

There have been many different suggestions of how to weight goodness of fit against model complexity.
Two in common use are the Akaike Information Critera (AIC) and the Bayesian Information Criteria
(BIC)~\cite{schwarz}. We tried both, and found neither to be particularly satisfactory, settling
instead on the Laplace approximation to the full Bayesian evidence. The evidence $p_X(s)$ for a model $X$
given data $s$ is given by the integral
\begin{equation}\label{marginal}
p_X(s) = \int p(s \vert \vec{\lambda},X) p(\vec{\lambda},X) d\vec{\lambda} \, .
\end{equation}
Computing this integral is extremely expensive for high dimension models, but the 
Laplace approximation provides the estimate: 
\begin{equation}
p_X(s) \simeq p(s \vert \vec{\lambda}_{\rm ML},X) \left( \frac{\Delta V_X}{V_X} \right) \, ,
\end{equation}
where $p(s \vert \vec{\lambda}_{\rm ML},X)$ is the maximum likelihood for the model, $V_X$ is the volume
of the model's parameter space, and $\Delta V_X$ is the volume of the uncertainty ellipsoid
(which we estimate using a Fisher Information Matrix). In general, adding another source template to 
the model will increase the likelihood, however, the $\Delta V_X/V_X$ term penalizes larger models
and serves as a built in Occam factor.

\begin{figure}[h]
\includegraphics[angle=270,width=0.5\textwidth]{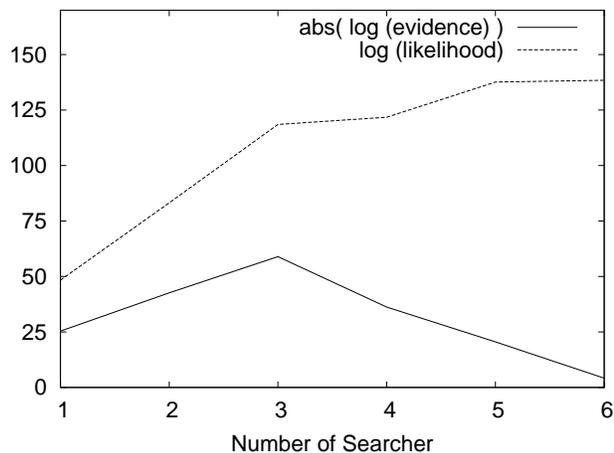}
\caption{\label{stopping_criteria_example_evidence_figure}Plot of the magnitude of the
logarithms of the likelihood and evidence for increasing numbers of searchers, searching
a data snippet with sources whose parameters are listed in
Table~\ref{stopping_criteria_example_parameters_table} injected into it.}
\end{figure}

As an example we will look at a data snippet containing $4$ sources. The source parameters and SNRs are
shown in Table~\ref{stopping_criteria_example_parameters_table}. There is one dim source which is not
expected to be recovered (see Section~\ref{false_negatives} for more details).
Figure~\ref{stopping_criteria_example_evidence_figure} plots the logarithms of the evidence and likelihood
for models of increasing size. One can see that the evidence is peaked at the model with 3 source
templates, while the likelihood continues to climb as the model dimension is increased.
This tells us that the data favors a model with $3$ sources over one with $4$.

\begin{table}[t]
\caption{\label{stopping_criteria_example_parameters_table}Source Parameter for the $4$ sources in the data snippet}
\begin{tabular}{|l|c|c|c|c|c|c|c|}
\hline
 SNR & $A$ ($10^{-24}$) & $f$ (mHz) & $\theta$ & $\phi$ & $\psi$ & $\iota$ & $\varphi_0$ \\
 \hline 
1.23 & 2.05 & 3.001514406 & 1.259 & 4.012 & 0.759 & 1.183 & 2.551 \\
7.67 & 10.8 & 3.000315843 & 2.437 & 2.753 & 2.484 & 2.173 & 2.550 \\
9.65 & 12.1 & 3.000454748 & 2.198 & 0.422 & 2.880 & 2.263 & 2.991 \\
10.76 & 9.27 & 3.001985584 & 1.336 & 5.863 & 0.931 & 2.805 & 4.048 \\
\hline  
\end{tabular}
\label{tab1}
\end{table}

An alternative approach to model selection is to use the Reverse Jump MCMC algorithm, which allows
for transitions between models of different dimension. The fraction of the time the chain spends
exploring each model is used as a measure of the relative evidence for the different models. We
plan to use the RJMCMC approach in future versions of the algorithm.

One limitation of the way we have formulated our Bayesian model selection is that the evidence
for a zero source model is ill defined, so we cannot compare models with 0 and 1 source templates.
This limitation can be removed if we expand our models to include the instrument noise parameters,
which we plan to do in future versions of the algorithm.

For those that favor the Frequentist approach to data analysis we provide some rough estimates of
the false alarm and false dismissal rates in the following subsections.

\subsection{False Negatives}\label{false_negatives}

\begin{figure}[h]
\includegraphics[angle=270,width=0.5\textwidth]{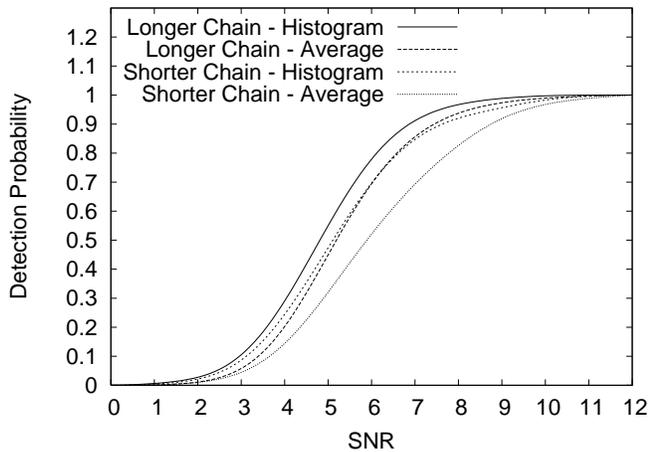}
\caption{\label{false_negatives_plot}Probability of detection of a source using the BAM algorithm as a function of source SNR.}
\end{figure}

Here we study the detection rate for the BAM algorithm as a function of the SNR of an isolated source.
To this end we performed searches of data snippets between $3 {\rm mHz}$ and $3.031688 {\rm mHz}$
(a $1000 f_m$ segment of the LISA band), each containing a single source.
A set of parameters for $100$ sources were chosen at random. Keeping all other parameters constant,
the amplitudes were varied to increase the SNR. Simulated data sets for each of the 100 sources were
created at different SNR levels. The BAM algorithm was then applied to the simulated data using
short chains of $15,000$ steps ($10,000$ steps in the annealing phase and $5,000$ set in the sampling
phase of each run). The searches were also run using longer chains, consisting of $75,000$ steps
($50,000$ annealing/$25,000$ sampling). For each of two types of chains two analysis methods were used
to determined the parameter values. In one method the parameter values were determined by using
average values of the parameters in the chain from the sampling phase ({\it i.e.} Bayes estimates).
In the second method the parameter values were determined from the mode of the parameter histograms
from the sampling phase ({\it i.e.} MAP estimates). Figure~\ref{false_negatives_plot} shows the
detection probabilities for these searches based on the two analysis methods. The resulting parameter
set is called a detection when they deviate by less than $5-\sigma$ in each of the true source's
intrinsic parameters (for more discussion on this cut-off see Section~\ref{example_searches}).

As expected, the detection probability depends on the length of the chain, with the longer chains
giving higher detection rates. However, a single very long chain is not the best way to ensure
detection. Consider the search for sources with ${\rm SNR}=5$ using the MAP parameter estimates.
The detection rate for the shorter chains were $0.45$, while for the the longer chain the rate was
$0.57$. However, if the search for a single ${\rm SNR}=5$ source is repeated multiple times, the
detection rate also comes out at $\sim 0.45$. Thus, if the short chain search is run twice and
the results from the two runs are combined, the detection probability improves to 
$\sim 1-0.55^2 = 0.7$, at a total cost of 30,000 steps, which is still less than half the cost
of the long chain searches.

Of the two methods for determining the source parameters, the MAP performed better than the Bayes estimate
in the low SNR region. On the other hand, the MAP is harder to compute, especially when there are a
large number of sources, so the current implementation of the full scale BAM algorithm still uses
Bayes estimates. This will be corrected in future versions.

\subsection{False Positives}

We now turn our attention to computing the false alarm rate by searching data streams that contain only
instrument noise. We employed two approaches: in the first we perform multiple searches with differing
noise realizations and in the second we performed an extended search of one noise realization.

\begin{figure}[h]
\includegraphics[angle=270,width=0.5\textwidth]{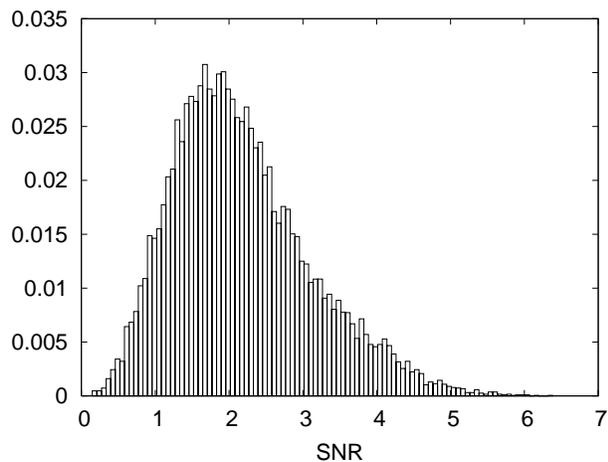}
\caption{\label{false_positives_SNR_1}Histogram of the SNRs for searches on source-free noise in a $100 f_m$ band starting at $3 {\rm mHz}$.}
\end{figure}

For the first test we performed $20,000$ searches using a single search template in the same frequency band
we used to study the false negatives.
Figure~\ref{false_positives_SNR_1} shows a histogram of the SNRs for the finishing points of the
searches. This can be used to give an idea as to the level where false positive begin to become a
concern. In this frequency range, the histogram tells us that more than $99 \%$ of the searches ended
on parameters leading to a SNR less than $5$. So in accepting any result from a search with SNR
above $5$ in this regime there is roughly a $1 \%$ chance of accepting a false positive, with
the probability dropping precipitously for searchers returning higher SNRs. We repeated the analysis
at different frequencies and found the false positive level to be much the same across the
LISA frequency band.

\begin{figure}[h]
\includegraphics[angle=270,width=0.5\textwidth]{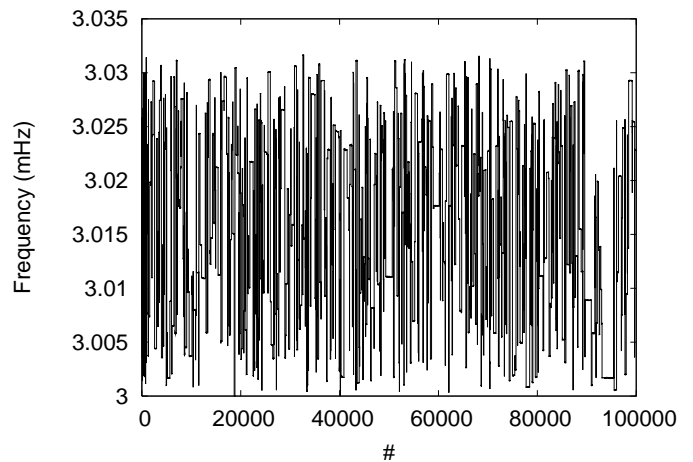}
\caption{\label{frequency_steps}Plot of the frequency steps in a chain searching source-free noise in a $1000 f_m$ band starting at $3 {\rm mHz}$.}
\end{figure}

\begin{figure}[h]
\includegraphics[angle=270,width=0.5\textwidth]{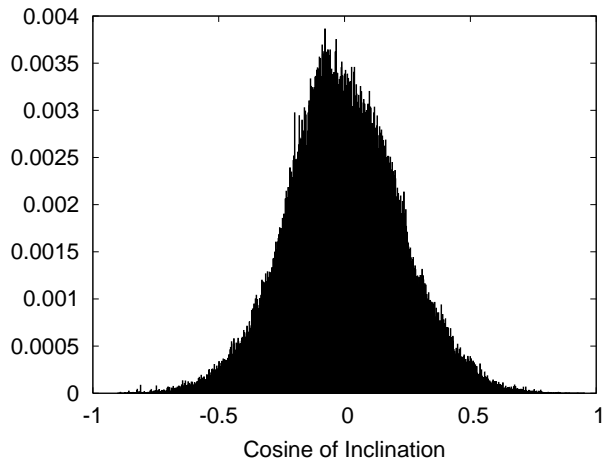}
\caption{\label{cos_inclination_histogram}Histogram of the cosine of the inclination angle for a chain searching source-free noise in a $1000 f_m$ band starting at $3 {\rm mHz}$.}
\end{figure}

\begin{figure}[h]
\includegraphics[angle=270,width=0.5\textwidth]{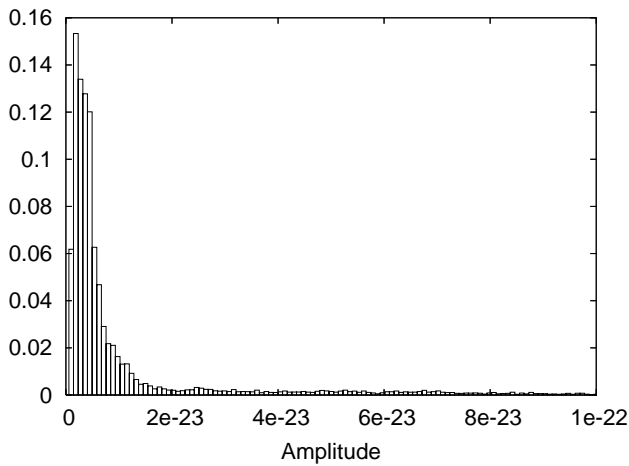}
\caption{\label{Amplitude_histogram}Histogram of the amplitude parameter for a chain searching source-free noise in a $1000 f_m$ band starting at $3 {\rm mHz}$.}
\end{figure}

The second test used the same $1000 f_m$ band at $3 {\rm mHz}$, but now a single long chain of
one million steps was performed. Figure~\ref{frequency_steps} shows the frequency parameter over
$100,000$ steps in the chain. One notices that the chain does not lock in on any particular frequency,
rather it continually wanders about the entire search region. While this is common, it is not always
the case, as instrument noise can sometimes mimic a monochromatic source well enough to slow or even
stop such exploratory movement in a chain. Two other reliable indicators of a false positive are the
amplitude and the inclination angle of the binary system. Figures~\ref{cos_inclination_histogram}
and \ref{Amplitude_histogram} show their respective histograms. The cosine of the inclination angle is
peaked around zero, and the amplitude is peaked at a level that gives templates whose spectral density
has the same magnitude as the noise. The reason for these preferences is that they allow the template
to optimally match the noise in the two LISA data channels. An inclination angle of $\pi/2$ gives
equal weight to the two channels, which allows the template to match the noise level in both channels
equally well.

To summarize, if the BAM algorithm recovers a source with ${\rm SNR} \geq 5$ it is unlikely to be a
false positive. Moreover, the chains show some very characteristic features when the templates are
just matching noise, and these features can be used as a diagnostic to exclude false positives.
Lastly, one can always run the search multiple times, and if the same same set of parameters are
recovered over and over again it is less likely that we have found a false positive.

\section{Example Searches}\label{example_searches}

In this section we show the results of several types of searches performed with the BAM algorithm.
While results for single source searches are easy to describe, when the search is for hundreds of sources,
and there are hundreds of successes, the numbers are much harder to present in a condensed manner. To that
end, we will present multiple search cases using plots like those in Figure~\ref{hierarchical_example},
where the frequency values are shown, and we will list the deviations of the recovered intrinsic parameters
scaled by the uncertainties given by the Fisher Information Matrix for the sources. Since we are
mostly interested in the search phase of the algorithm the post-search MCMC runs we chosen to be rather
short, so the recovered posteriors are a little ragged. With this in mind we set our cut-off criteria
for ``finding'' a source at a deviation of $5-\sigma$ from any one of the true source's intrinsic
parameters. In practice this cut-off was fairly robust as any template that did not have all parameters
within a few $\sigma$ of the source typically had one or more parameters that were tens to
hundreds of $\sigma$ out. We also imposed was a SNR minimum of $5$ in addition to the Bayesian evidence
criteria. This was perhaps stricter than
necessary to keep down the possibility of a false positive (and in fact did lead to one actual
detection being discounted), but with that cut-off there were no instances of false positives in
any of the searches presented here.

\subsection{Searches of the Mock LISA Data Challenge Training Sets}

The Mock LISA Data Challenge (MLDC) consists of several types of challenges for the LISA data
analysis community to test search algorithms using simulated LISA data. The first round in the
MLDC consists of challenges for three source types: galactic binaries, supermassive black holes,
and extreme mass ratio inspirals. For this work we will focus on the challenges dealing with
galactic binaries. Provided with each challenge are two data sets, one is a blind test where the
source parameters are unknown, while the other is an open test where the source parameters are
provided so that one may synchronize conventions. In this paper we will only be discussing results
from the open data sets, as the MLDC Taskforce has asked that all results for the blind challenges
be embargoed until December 2006.

\subsubsection{Single binary searches}

The first challenges consist of single binary systems injected into a LISA data stream with
instrumental noise. In the Challenge 1.1.1 there are three tests. The first is a monochromatic
binary with frequency, $f = 1.0 \pm 0.1 {\rm mHz}$, the second has a frequency of
$f = 3.0 \pm 0.1 {\rm mHz}$, and the third a frequency of $f = 10.0 \pm 1.0 mHz$. While the BAM
algorithm is designed to handle multiple source searches these single source challenges provide
a means to test our conventions as well check for any modeling errors. One source
of modeling error is that we used a low frequency detector response model, which restricts us to
searching for signals below 7 mHz. Thus we have only performed searches on the first two of these
single source challenges.

In the $1 {\rm mHz}$ case the true source parameters and the results from the BAM algorithm are given
in Table~\ref{MLDC_1.1.1a_Training_Set}. For this work all uncertainties will be calculated using a
Fisher Information Matrix at the parameter values given by the chain. A longer run of the data chains
after burn-in could also provide a means to calculate these uncertainties, but as was shown in~\cite{MCMC}
for the intrinsic parameters these will be a very good match to those from the Fisher calculation.
Here the deviations (or discrepancies) between the true parameters value of the three intrinsic
parameters and the values determined through the search are less than $1 \sigma$: ${\Delta f} \equiv (f_{true} - f_{search}) = -0.7530 \sigma_f$, ${\Delta \theta} = -0.5563 \sigma_\theta$, and
${\Delta \phi} = -0.20498 \sigma_\phi$.

\begin{table*}[t]
\caption{\label{MLDC_1.1.1a_Training_Set}Results of a search of the MLDC Training Data Set 1.1.1a}
\begin{tabular}{|l|c|c|c|c|c|c|c|}
\hline
  & $A$ ($10^{-22}$) & $f$ (mHz) & $\theta$ & $\phi$ & $\psi$ & $\iota$ & $\varphi_0$ \\
 \hline 
True Parameters & 1.789229908 & 0.9930348535 & 0.4741143268 & 5.19921 & 3.975816 & 0.1793956 & 5.781211 \\
Recovered Parameters & 2.364 & 0.993034139 & 0.4575 & 5.196 & 4.475 & 0.7364 & 4.832 \\
Parameter Uncertainties & 0.3876 & 9.493e-07 &  0.02983 & 0.01754 & 0.3126 & 0.2095 & 0.6234 \\
\hline  
\end{tabular}
\label{tab1}
\end{table*}

For the $3 {\rm mHz}$ case the search returned mean parameter values that were also discrepant
from the true source parameters by less than $1 \sigma$ (${\Delta f} = 0.2164 \sigma_f$,
${\Delta \theta} = -0.09466\sigma_\theta$, ${\Delta \phi} = -0.7860 \sigma_\phi$).

In these two searches the algorithm is performed exactly as expected. This suggests that our current
implementation of the algorithm is free of any significant systematic errors in either waveform generation
or calculation of the likelihood values, as the MLDC data was created with an independently developed code.

\subsubsection{Low Source Confusion}

Challenge 1.1.4 for the MLDC is a test for algorithms in the low source confusion regime, where the source
density is $\sim 1$ source per $10 f_m$. Results for a search of the training data set are presented here.

Using only the frequency range given in the challenge ($3.000 {\rm mHz} <$ f $< 3.015 {\rm mHz}$), the BAM
was run on the data stream (an approximate range of the number of source was given in the challenge, and
indeed the exact number in the training data is known to be $50$, but that information was not used in
directing the algorithm). The log evidence was used as the stopping criteria for each search region.
The frequency range of the data snippet was divided into $20$ search regions such that each acceptance
window was $0.75 \mu{\rm Hz}$ in width, and the width of the wings were $0.1$ times the bandwidth of
a typical source with a frequency in the search region.

\begin{figure}[h]
\includegraphics[angle=270,width=0.5\textwidth]{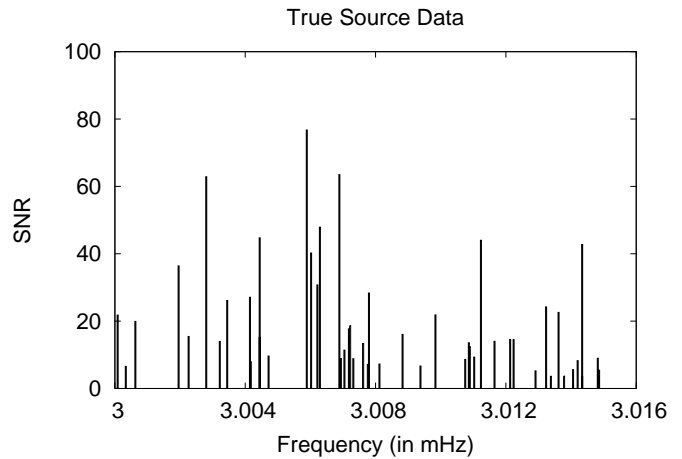}
\caption{\label{Training_data_1.1.4_source_data}Plot of the true parameter frequencies and their respective SNRs for the sources injected into a LISA data stream for Training data set 1.1.4 of the Mock LISA Data Challenge.}
\end{figure}

\begin{figure}[h]
\includegraphics[angle=270,width=0.5\textwidth]{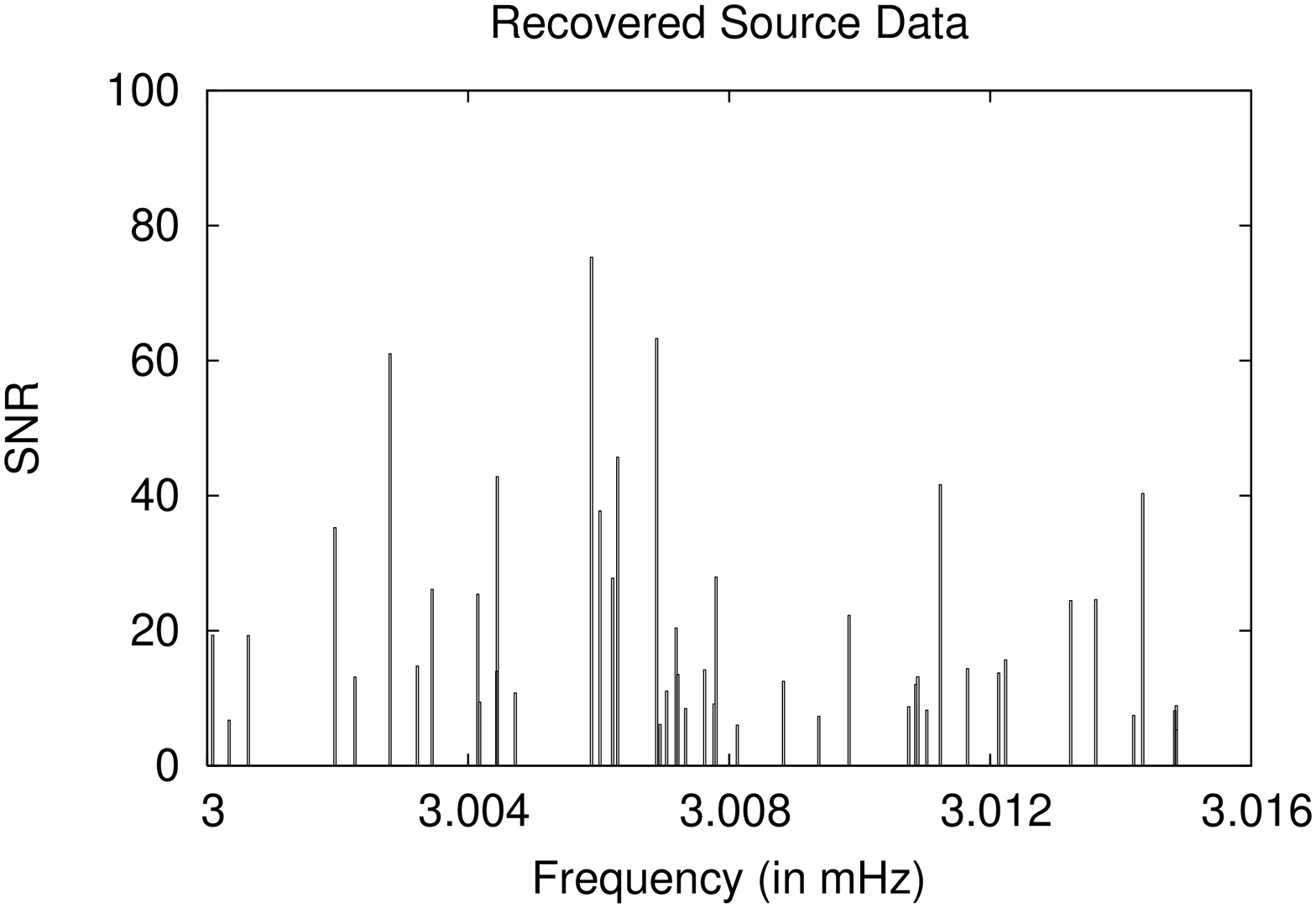}
\caption{\label{Training_data_1.1.4_recovered_data}Plot of the recovered parameter frequencies and their respective SNRs for the sources recovered by the BAM algorithm searching Training data set 1.1.4 of the Mock LISA Data Challenge.}
\end{figure}

\begin{figure}[h]
\includegraphics[angle=270,width=0.5\textwidth]{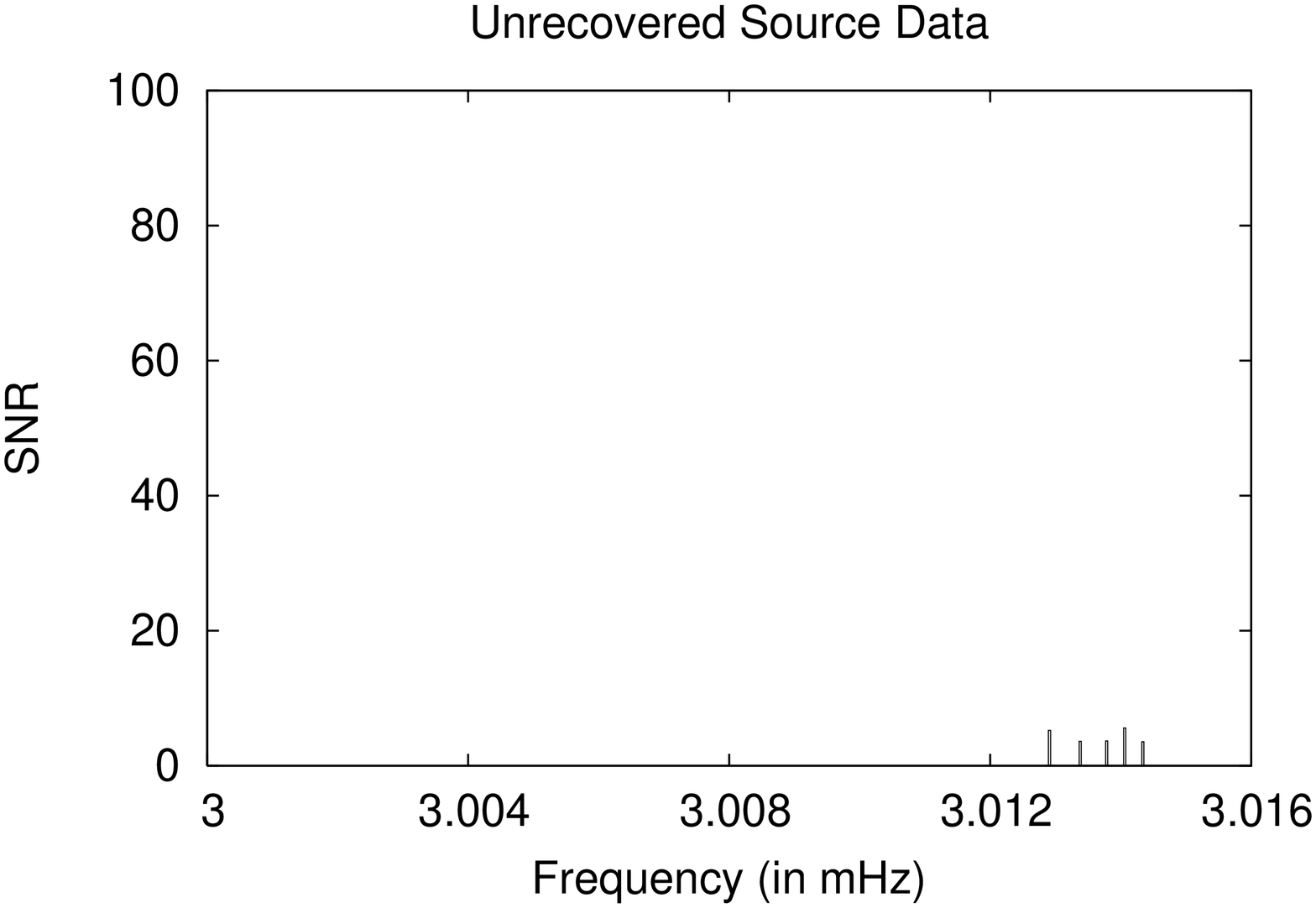}
\caption{\label{Training_data_1.1.4_unrecovered_data}Plot of the true parameter frequencies and their respective SNRs for the sources that were not recovered by the BAM algorithm searching Training data set 1.1.4 of the Mock LISA Data Challenge.}
\end{figure}

Figure~\ref{Training_data_1.1.4_source_data} shows a plot of the locations in frequency of the $50$
individual sources in the data snippet. The heights of the bars show the SNR of each source, while the
width of the bars is $1 f_m$. Results of the search are similarly expressed in
Figure~\ref{Training_data_1.1.4_recovered_data}. As these two plots are very similar
Figure~\ref{Training_data_1.1.4_unrecovered_data} has been provided to highlight the
difference between the them. 

\begin{figure}[h]
\includegraphics[angle=270,width=0.5\textwidth]{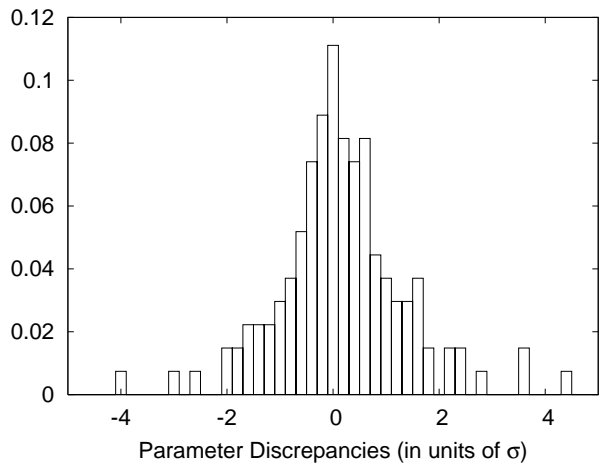}
\caption{\label{Training_data_1.1.4_sigma_histogram}Histogram of the discrepancies between the true intrinsic source parameters and the intrinsic parameters recovered by the BAM algorithm searching Training data set 1.1.4 of the Mock LISA Data Challenge. Differences are given in units of the parameter variances.}
\end{figure}

The five sources shown in Figure~\ref{Training_data_1.1.4_unrecovered_data} represent the false
negatives of this particular search. Three of these sources had SNRs $< 5$ and thus were not expected
to be recovered given the results of subsection~\ref{false_negatives}. The remaining two unrecovered
sources had SNRs $< 5.6$. Figure~\ref{Training_data_1.1.4_sigma_histogram} shows how well the search
algorithm fit the true source parameters. It is a histogram of the differences in the $135$ intrinsic
parameters of the recovered sources in units of their respective variances (as calculated via a
Fisher Information Matrix located at the recovered parameter values). Nearly $92 \%$ of the parameters
recovered by the algorithm differed from their true values by less than $2 \sigma.$ This fit might
be improved some with a so-called 'finisher' step, which will be briefly discussed in more detail in
the following subsection.

\subsubsection{Strong Source Confusion}

Challenge 1.1.5 for the MLDC is a test for algorithms in the high source confusion regime, where the source
density is $\sim 1$ source per $\sim 2.5 f_m$. Results for a search of the training data set are presented here.

The BAM algorithm was run on the data streams using the frequency range given in the challenge
($2.9985 {\rm mHz} < f < 3.0015 {\rm mHz}$). Again, the approximate range of the number of sources
that was given in the challenge was not used in directing the algorithm. The log evidence was used as
the stopping criteria for each search region. The frequency range of the data snippet was divided into
$10$ search regions such that each acceptance window was $0.3 \mu{\rm Hz}$ in width, and the width
of the wings were $0.5$ times the bandwidth of a typical source in the search region,
${\rm BW} = 28 f_m $. The wing size of this search is considerably larger than was used in the
previous example. Initial runs on this data set with smaller wings showed signs of slamming, so
the wing size was increased.

\begin{figure}[ht]
\includegraphics[angle=270,width=0.5\textwidth]{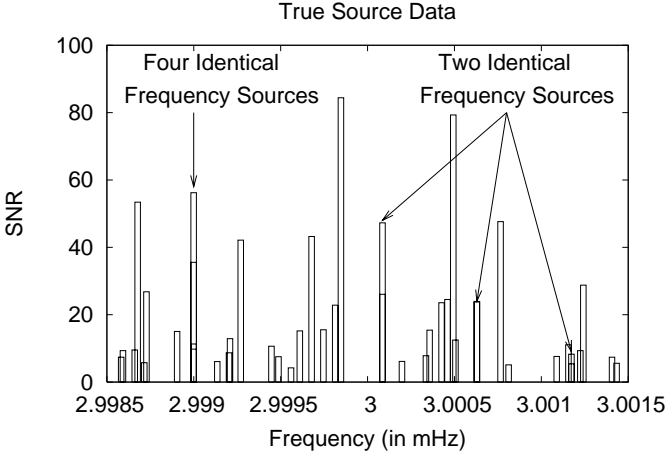}
\caption{\label{source_data_figure}Plot of the true parameter frequencies and their respective SNRs for the sources injected into a LISA data stream for Training data set 1.1.5 of the Mock LISA Data Challenge.}
\end{figure}

Figure~\ref{source_data_figure} shows a plot of the locations in frequency of the individual sources
in the data snippet. The heights of the bars show the SNR of each source, while the width of the
bars is $1 f_m$. The density of sources is even higher than first appears in the plot, however, since
four sources share an identical frequency ($2.998999384 {\rm mHz}$) as do three other pairs of
sources ($3.000085802$, $3.000629082$, and $3.001173008 {\rm mHz}$). 

\begin{figure}[h]
\includegraphics[angle=270,width=0.5\textwidth]{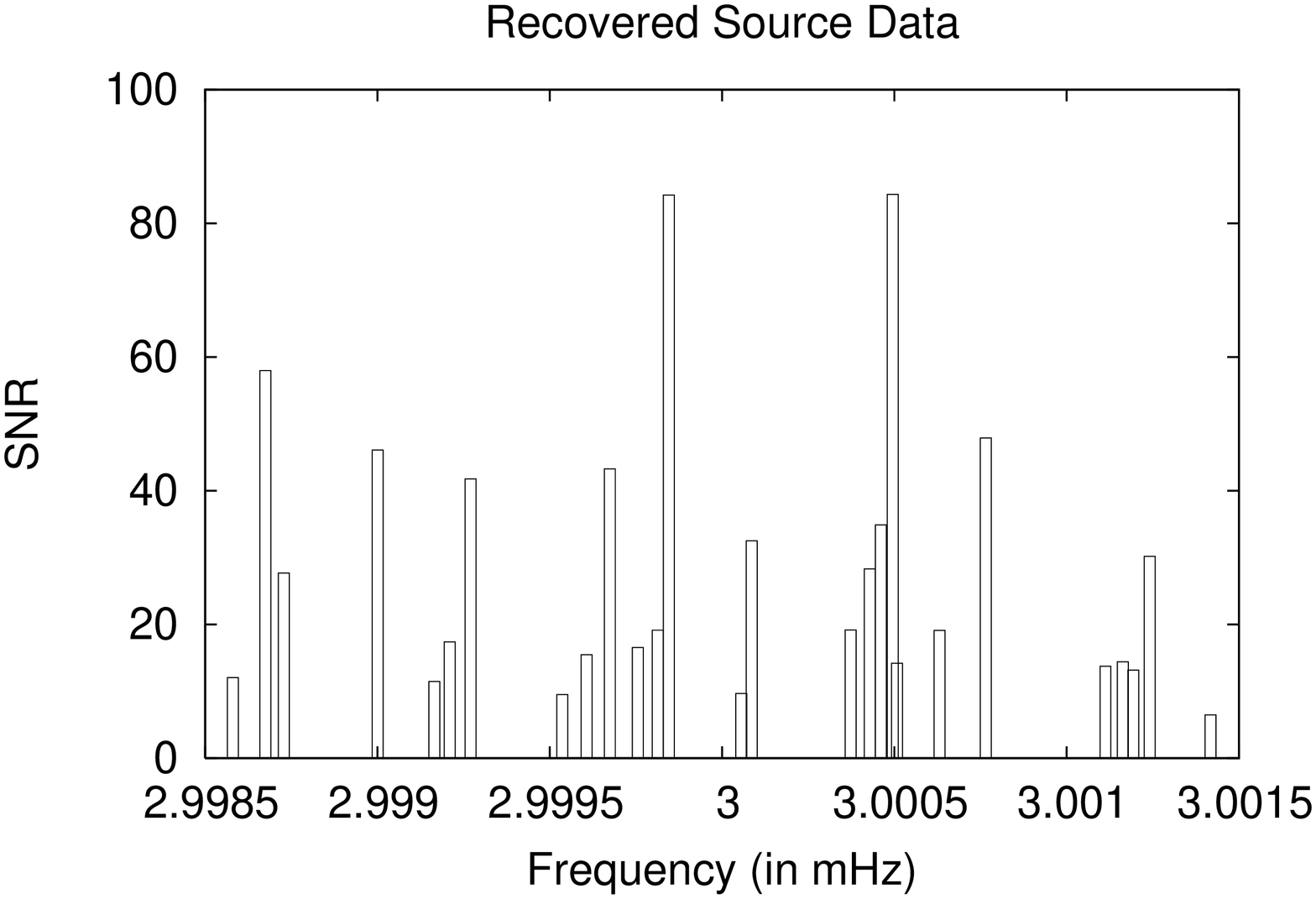}
\caption{\label{Training_data_1.1.5_recovered_data}Plot of the recovered parameter frequencies and their respective SNRs for the sources recovered by the BAM algorithm searching Training data set 1.1.5 of the Mock LISA Data Challenge.}
\end{figure}

\begin{figure}[h]
\includegraphics[angle=270,width=0.5\textwidth]{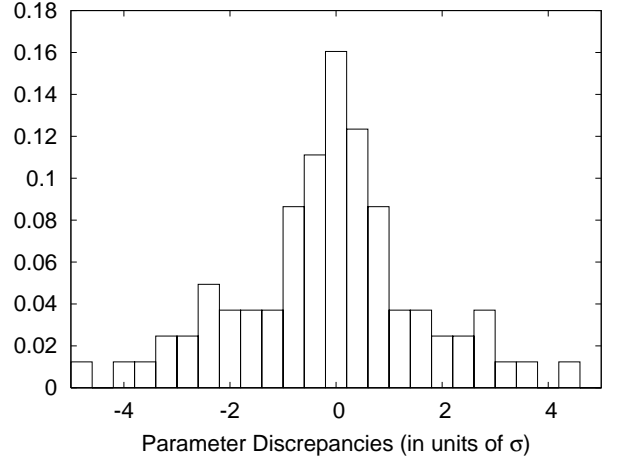}
\caption{\label{Training_data_1.1.5_sigma_histogram}Histogram of the discrepancies between the true intrinsic source parameters and the intrinsic parameters recovered by the BAM algorithm searching Training data set 1.1.5 of the Mock LISA Data Challenge. Differences are given in units of the parameter variances.}
\end{figure}

Results of the search are displayed in Figure~\ref{Training_data_1.1.5_recovered_data}. As can be
seen in the plot, the extremely high density of sources prevent the algorithm from recovering all
of the sources. Of the $44$ sources injected into the data stream, $27$ were recovered, which
corresponds to a recovered source density of 1 per 4 frequency bins.
Figure~\ref{Training_data_1.1.5_sigma_histogram} shows a histogram of the differences in the $81$
intrinsic parameters of the recovered sources in units of their respective variances
(as calculated via a Fisher Information Matrix located at the recovered parameter values).
The spread of the discrepancies for the recovered sources is larger than those for the case of
low source confusion shown in Figure~\ref{Training_data_1.1.4_sigma_histogram}. Just
over $76 \%$ percent of the parameters recovered by the algorithm differed from their true
values by less than $2 \sigma.$ This departure from the Fisher matrix predictions is due to
the additional confusion noise from unrecovered sources.

\begin{figure}[h]
\includegraphics[angle=270,width=0.5\textwidth]{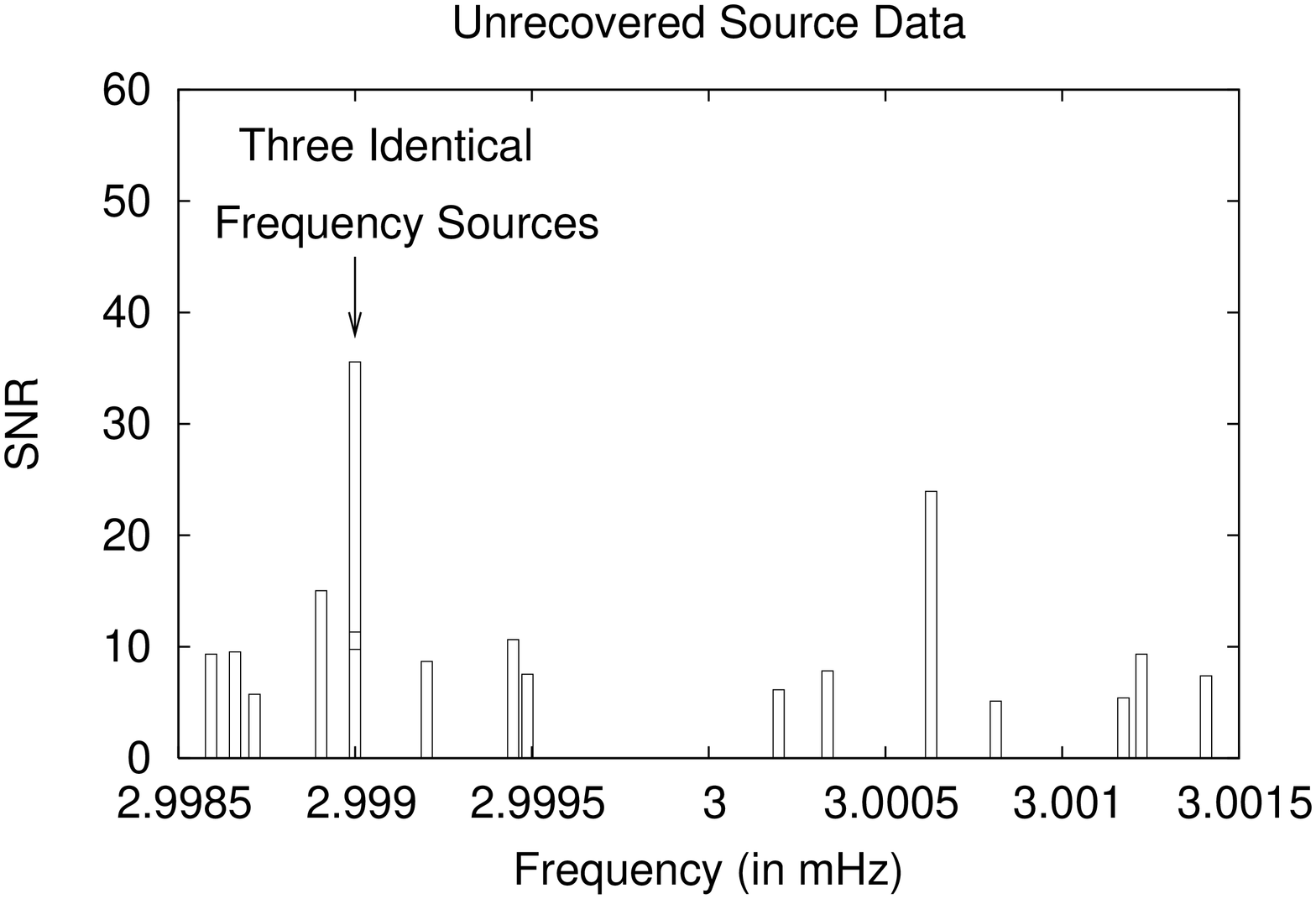}
\caption{\label{Training_data_1.1.5_unrecovered_data}Plot of the true parameter frequencies and their respective SNRs for the sources that were not recovered by the BAM algorithm searching Training data set 1.1.5 of the Mock LISA Data Challenge.}
\end{figure}

Figure~\ref{Training_data_1.1.5_unrecovered_data} shows the sources that were not recovered by
algorithm. Of the $17$ unrecovered sources, $13$ had a nearest neighbor that was within $1 f_m$,
including $5$ of the unrecovered sources that shared an identical frequency with at least one other
source.  Sources that are close in frequency are much more likely to be highly correlated than those
that are well separated (e.g. the brightest of the sources at $f = 2.998999384 {\rm mHz}$ anti-correlates
with two of the other sources at that frequency at the level of $-0.81$ and $-0.67$). This high density
and corresponding high levels of correlation introduces two problems for the current implementation
of the BAM algorithm. First, is that analysis of the chains was done using the mean values of a
particular search template. With nearby sources, the individual searchers can jump between the sources
and the calculated mean value is a weighted mean of the two, or more, close sources (weighted by the
number of steps in the chain spent at each source). In the next implementation of the algorithm we
combine the all the parameter chains of a given type into a single histogram and use standard spectral
line fitting software to fit the combined PDF by multiple Gaussians. Second, the current implementation
of the algorithm includes a ``blast'' proposal distribution that separates highly anti-correlated sources
($\kappa < -0.9$). This proposal was included to lessen the effect of slamming (by performing a uniform
draw on one of the two anti-correlated searchers). With the inclusion of the wing noise and returning
to a $7$ parameter search (per template), this proposal will most likely not be needed. This should allow
the search templates to spend more time in areas with highly anti-correlated sources.

Lastly, the fit could be improved using a 'finisher' step in the analysis process. While the full details
of such a finisher are beyond the scope of this paper, one can think of it as continuing the search
algorithm using proposals specific to providing efficient mixing of the chain (such as a drawing from
a multivariate gaussian distribution) that will allow for searchers to work through the issues created
by the high levels of correlation and reach the posterior distribution. Also, in this step sources can
be introduced to the fit given by the BAM algorithm to provide a better fit.

\subsection{A Large $N$ Search for Resolvable Binaries}\label{NYZspin}

In this subsection we will discuss the results of a search for sources in a $1000 f_m$ data snippet
at $3 {\rm mHz}$. These sources were chosen from a galactic model described by
Nelemans {\it et al}~\cite{gils}. For this realization there are $281$ sources in the data snippet.

\begin{figure}[h]
\includegraphics[angle=270,width=0.5\textwidth]{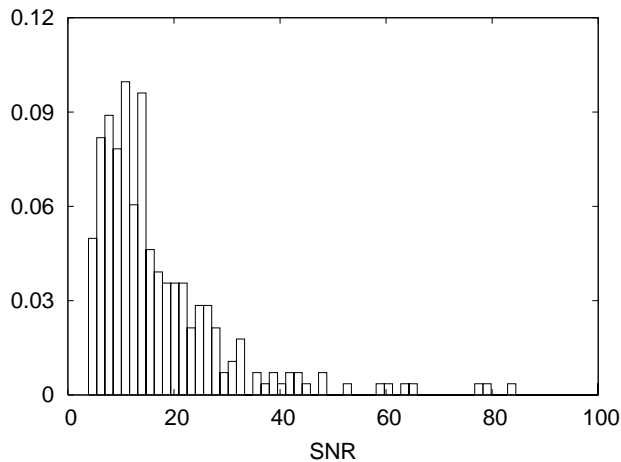}
\caption{\label{1000fm_SNR_histogram}Partial histogram of the SNRs for the true source parameters
of a $1000 f_m$ data snippet containing $281$ sources with a $3$ year time of observation.}
\end{figure}

Unlike the previous searches, the observation time is $3$ years. This provides a test of the algorithm
for multi-year observation times, and models a search through a non-trivial section of the galactic
background (nearly $1\%$ of the overall frequency range, and $>1\%$ of the expected resolvable sources).
With a three year observation time all but four of the sources have a SNR $> 5$.
Figure~\ref{1000fm_SNR_histogram} shows a histogram of the SNRs for sources below $100$
(another $9$ sources have SNRs $> 100$).

\begin{figure}[h]
\includegraphics[angle=270,width=0.5\textwidth]{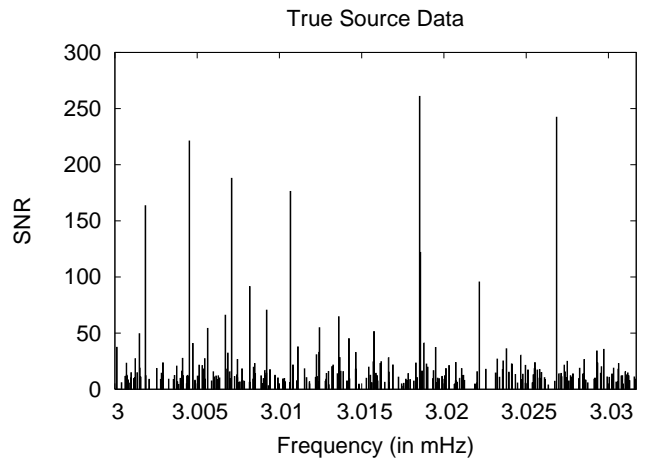}
\caption{\label{1000fm_source_data}Plot of the true parameter frequencies and their respective
SNRs for the sources injected into a LISA data stream for $281$ sources drawn from a galactic
distribution with a $3$ year time of observation.}
\end{figure}

\begin{figure}[h]
\includegraphics[angle=270,width=0.5\textwidth]{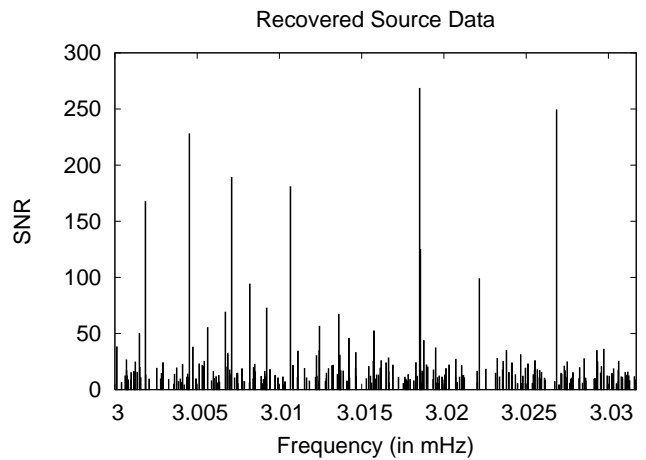}
\caption{\label{1000fm_recovered_data}Plot of the $270$ recovered parameter frequencies and their
respective SNRs for the sources recovered by the BAM algorithm searching a $1000 f_m$ data snippet
containing $281$ sources with a $3$ year time of observation.}
\end{figure}

\begin{figure}[h]
\includegraphics[angle=270,width=0.5\textwidth]{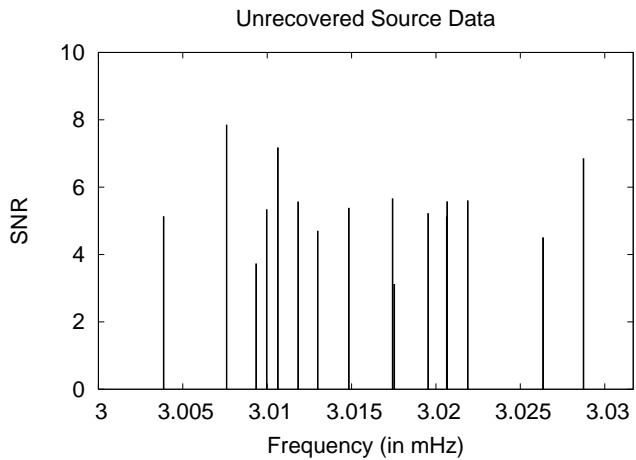}
\caption{\label{1000fm_unrecovered_data}Plot of the $16$ true parameter frequencies and their
respective SNRs for the sources that were not recovered by the BAM algorithm searching a
$1000 f_m$ data snippet containing $281$ sources with a $3$ year time of observation.}
\end{figure}

Figure~\ref{1000fm_source_data} shows a plot of the locations in frequency of the individual sources
in the $1000 f_m$ data snippet. The heights of the bars show the SNR of each source, while the width
of the bars is $1 f_m$. The results of the search are shown in Figure~\ref{1000fm_recovered_data}.
As these two plots are very similar, Figure~\ref{1000fm_unrecovered_data} has been provided
(and re-scaled) to highlight the difference between the them. The search was able to find $265$ of
the $281$ sources, with all but $3$ of the unrecovered sources having a SNR $< 6$. One of these
``unrecovered'' sources was an instance of the SNR cut-off inadvertently omitting a detected source
(the source at f $= 3.02634519$ has a SNR $= 4.50$, and it was recovered with a SNR $= 4.62$). It is
interesting to note that these results are consistent with the prediction in Section~\ref{false_negatives}
regarding the rate of false negatives. This suggests that most of the $12$ sources that were not recovered
and have a SNR $> 5$ might be found with repeated searches.

\begin{figure}[h]
\includegraphics[angle=270,width=0.5\textwidth]{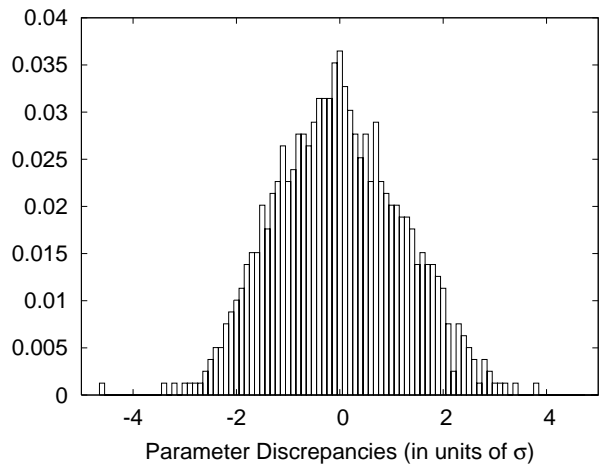}
\caption{\label{1000fm_sigma_histogram}Histogram of the discrepancies between the true
intrinsic source parameters and the intrinsic parameters recovered by the BAM algorithm
searching a $1000 f_m$ data snippet containing $281$ sources with a $3$ year time of
observation. Differences are given in units of the parameter variances.}
\end{figure}

Figure~\ref{1000fm_sigma_histogram} shows a histogram of the differences in the $795$ intrinsic
parameters of the recovered sources in units of their respective variances (as calculated via a
Fisher Information Matrix located at the recovered parameter values). Slightly more than $91 \%$
of the parameters recovered by the algorithm differed from their true values by less than $2 \sigma$.

\section{Conclusion}\label{conclusion}

We have developed and tested an algorithm that is capable of locating and characterizing galactic
binaries across the entire LISA band. In regions of strong source confusion we found that the
algorithm could recover 1 source every 4 frequency bins. We found that
the algorithm performs very well on snippets taken from a full galactic foreground model, and
since the BAM algorithm develops a global solution by sewing together searches over subsets
of the LISA data, completing the full analysis of the galactic simulation is just a matter of computer time.

While the current algorithm is very effective, we identified many improvements and extensions that
are now being implemented. The waveform modeling has been updated to include the
full LISA response and frequency evolution, and we have reverted to performing full parameter searches
due to the computational cost incurred by the multi-source F-statistic. Work is currently in progress
to extend the search to include parameters that describe the noise in each data channel. This
extension will be particularly important below 3 mHz as the effective noise level will be set by
unresolved sources, so we will not know in advance what weighting to use in the inner products.
Our current stopping criteria using the Laplace approximation to the Bayes evidence could be
eliminated if we switch to a transdimensional Reverse-Jump MCMC method~\cite{rj}. The analysis of
the post-search chains can also be improved using spectral line fitting techniques. 

Even with our current algorithm we estimate that it would take less than two weeks to process
a full galactic foreground on a 3 GHz, 128 node cluster. With the modifications we are implementing
we expect both the speed and fidelity of the algorithm to be much improved. We will have an
opportunity to test the updated BAM algorithm on a full galactic simulation in the second round
of Mock LISA Data Challenges, which are set for release in December 2006.

\section*{Acknowledgments}
This work was supported at MSU by NASA Grant NNG05GI69G. A portion of the research described in this paper was
carried out at the Jet Propulsion Laboratory, California  Institute of Technology, under a contract
with the National  Aeronautics and Space Administration.


\begin{thebibliography}{99}

\bibitem{lppa} P. Bender {\it et al.}, {\em LISA Pre-Phase A Report}, (1998).
\bibitem{evans} C. R. Evans, I. Iben \& L. Smarr, ApJ {\bf 323}, 129 (1987).
\bibitem{lip} V. M. Lipunov, K. A. Postnov \& M. E. Prokhorov, A\&A {\bf 176}, L1 (1987).
\bibitem{hils} D. Hils, P. L. Bender \& R. F. Webbink, ApJ {\bf 360}, 75 (1990).
\bibitem{hils2} D. Hils \& P. L. Bender, ApJ {\bf 537}, 334 (2000).
\bibitem{gils} G. Nelemans, L. R. Yungelson \& S. F. Portegies Zwart, A\&A {\bf 375}, 890 (2001).
\bibitem{seth} S. Timpano, L. J. Rubbo \& N. J. Cornish, Phys. Rev. D{\bf 73} 122001 (2006).
\bibitem{vech} A. Vecchio, Phys. Rev. D{\bf 70}, 042001 (2004).
\bibitem{lang} R. N. Lang, S. A. Hughes, gr-qc/0608062 (2006).
\bibitem{bbw} E. Berti, A. Buonanno \& C. M. Will, Class. Quant. Grav. {\bf 22} S943 (2005).
\bibitem{rw} K. J. Rhook \& J.S. B. Wyithe, Mon. Not. Roy. Astron. Soc. {\bf 361}, 1145 (2005). 
\bibitem{kz} S. M. Koushiappas \& A. R. Zentner, Astrophys. J. {\bf 639} 7,(2006).
\bibitem{cutbar} L. Barack \& C. Cutler, Phys. Rev. D{\bf 69}, 082005 (2004).
\bibitem{emri} J. R. Gair, L. Barack, T. Creighton, C. Cutler, S. L. Larson, E. S. Phinney \& M. Vallisneri, Class. Quant. Grav. {\bf 21}, S1595 (2004).
\bibitem{drasco} S. Drasco, gr-qc/0604115 (2006).
\bibitem{drasco_hughes} S. Drasco \& S. Hughes, gr-qc/0509101 (2005)
\bibitem{MCMC} N.J. Cornish \& J. Crowder, Phys. Rev. D{\bf 72} 043005 (2005).
\bibitem{MCMC_ed_neil} N.J. Cornish \& E.K. Porter, Class. Quant. Grav. {\bf 23} S761 (2006).
\bibitem{vecchio_1} E.D.L. Wickham, A. Stroeer \& A. Vecchio, Class. Quant. Grav. {\bf 23} S819 (2006).
\bibitem{MCMC_ed_neil2} N.J. Cornish \& E.K. Porter, gr-qc/0605135 (2006).
\bibitem{vecchio_2} A. Stroeer, J. Gair \& A. Vecchio, gr-qc/0609010 (2006).
\bibitem{genetic} J. Crowder \& N.J. Cornish Phys.Rev. D {\bf 73} 063011 (2006).
\bibitem{gclean} N.J. Cornish \& S.L. Larson, Phys. Rev. D{\bf 67}, 103001 (2003).
\bibitem{slicedice}  L.J. Rubbo, N.J. Cornish \& R.W. Hellings, gr-qc/0608112 (2006).
\bibitem{curt_michele_duncan} C.J. Cutler, M. Vallisneri, \& D.A. Brown (in preparation).
\bibitem{browns} S. D. Mohanty \& R. K. Nayak, Phys. Rev. D{\bf 73}, 083006 (2006).
\bibitem{time_freq} J. Gair \& L. Wen, Class. Quant. Grav. {\bf 22} S445 \& S1359 (2005).
\bibitem{metro} N. Metropolis, A. W. Rosenbluth, M. N. Rosenbluth, A. H. Teller \& E. Teller, J. Chem. Phys. {\bf 21}, 1087 (1953).
\bibitem{haste} W. K. Hastings, Biometrics {\bf 57}, 97 (1970).
\bibitem{gamer} D. Gamerman, {\em  Markov Chain Monte Carlo: Stochastic Simulation of Bayesian Inference}, (Chapman \& Hall, London, 1997).
\bibitem{cm1} N. Christensen \& R. Meyer, Phys. Rev. D{\bf 58}, 082001 (1998)
\bibitem{christ} N. Christensen \& R. Meyer, Phys. Rev. D{\bf 64}, 022001 (2001); N. Christensen, R. Meyer \& A. Libson, Class. Quant. Grav.{\bf 21}, 317 (2004);
C. Roever, R. Meyer, \& N. Christensen, Class. Quant. Grav.{\bf 23}, 4895 (2006); J. Veitch, R. UmstŠtter, R. Meyer, N. Christensen \& G. Woan. Class. Quant. Grav.{\bf 22}, S995 (2005); R. UmstŠtter, R. Meyer, R.J. Dupuis, J. Veitch, G. Woan \& N. Christensen, Class. Quant. Grav.{\bf 21}, S1655 (2004).
\bibitem{woan} N. Christensen, R. J. Dupuis, G. Woan \& R. Meyer, Phys. Rev. D{\bf 70}, 022001 (2004);
R. Umstatter, R. Meyer, R. J. Dupuis, J. Veitch, G. Woan \& N. Christensen, gr-qc/0404025 (2004).
\bibitem{andrieu} Andrieu, C. and Doucet, A. (1999). Joint Bayesian model selection and estimation of noisy sinusoids via reversible jump MCMC. IEEE Trans. Signal Process. 47 2667--2676
\bibitem{woan2} R. Umstatter, N. Christensen, M. Hendry, R. Meyer, V. Simha, J. Veitch, S. Viegland \& G. Woan, gr-qc/0503121 (2005).
\bibitem{sam} L. S. Finn, Phys. Rev. D {\bf 46} 5236 (1992).
\bibitem{mcmc_hist} C. Andrieu, N. De Freitas, A. Doucet \& M. Jordan, Machine Learning {\bf 50}, 5 (2003).
\bibitem{gilks} {\em Markov Chain Monte Carlo in Practice}, Eds. W. R. Gilks, S. Richardson \& D. J. Spiegelhalter, (Chapman \& Hall, London, 1996).
\bibitem{fstat} P. Jaranowski, A. Krolak \& B. F. Schutz, Phys. Rev. D{\bf 58} 063001 (1998).
\bibitem{jensen} C. S. Jensen, A. Kong \& U. Kjaerulff, International Journal of Human Computer Studies, {\bf 42},
647 (1995).
\bibitem{cc} C. Cutler, Phys. Rev. D {\bf 57}, 7089 (1998).
\bibitem{cr} N. J. Cornish \& L. J. Rubbo, Phys. Rev. D{\bf 67}, 022001 (2003).
\bibitem{rigad} L. J. Rubbo, N. J. Cornish \& O. Poujade, Phys. Rev. D{\bf 69} 082003 (2004).
\bibitem{gemans}S. Geman \& D. Geman. "Stochastic Relaxation, Gibbs Distributions, and the Bayesian Restoration of Images". IEEE Transactions on Pattern Analysis and Machine Intelligence, 6:721-741, 1984
\bibitem{sterl} A. J. Farmer \& E. S. Phinney, Mon. Not. Roy. Astron. Soc. {\bf 346}, 1197 (2003).
\bibitem{bc}  L. Barack \& C. Cutler, Phys. Rev. D{\bf 70}, 122002 (2004).
\bibitem{schwarz} G. Schwarz, Ann. Stats. {\bf 5}, 461 (1978).
\bibitem{MLDC} K.A. Arnaud, S. Babak, J.G. Baker, M.J. Benacquista, N.J. Cornish, C. Cutler, S.L. Larson, B.S. Sathyaprakash, M. Vallisneri, A. Vecchio, J-Y. Vinet, gr-qc/0609105 (2006)
\bibitem{rj} P. J. Green \& A. Mira, Biometrika {\bf 88}, 1035 (2001).
\bibitem{Gelman}Andrew Gelman, John B. Carlin, Hal S. Stern, and Donald B. Rubin. Bayesian Data Analysis. London: Chapman and Hall. First edition, 1995. (See Chapter 11.)

\end{thebibliography}
\end{document}